\documentclass[acmsmall]{acmart}
\setcopyright{rightsretained}
\acmDOI{10.1145/3728943}
\acmYear{2025}
\acmJournal{PACMSE}
\acmVolume{2}
\acmNumber{ISSTA}
\acmArticle{ISSTA066}
\acmMonth{7}
\received{2024-10-31}
\received[accepted]{2025-03-31}

\usepackage{multirow}
\usepackage{tabularx}
\usepackage{tabulary}
\usepackage{rotating}
\usepackage{lscape}
\usepackage{enumitem}

\usepackage{listings}
\usepackage{xcolor}

\usepackage{caption}
\usepackage{subcaption}
\usepackage{ifthen}
\usepackage{xcolor}

\newcolumntype{L}[1]{>{\raggedright\let\newline\\\arraybackslash\hspace{0pt}}m{#1}}
\newcolumntype{C}[1]{>{\centering\let\newline\\\arraybackslash\hspace{0pt}}m{#1}}
\newcolumntype{R}[1]{>{\raggedleft\let\newline\\\arraybackslash\hspace{0pt}}m{#1}}
\definecolor{ao(english)}{rgb}{0.0, 0.5, 0.0}

\usepackage{tikz}
\newcommand{\blackcircle}[1]{%
    \tikz[baseline=(char.base)]{
        \node[shape=circle, fill=black, text=white, inner sep=1pt] (char) {#1};
    }%
}

\usepackage[most]{tcolorbox}
\newcommand{\rqbox}[1]{

\begin{tcolorbox}[tile, size=fbox, boxsep=2mm, boxrule=0pt, top=0pt, bottom=0pt,
borderline west={1mm}{0pt}{gray!50!white}, colback=gray!10!white]
#1
\end{tcolorbox}

}
\usepackage{fontawesome5}

\usepackage{color}
\usepackage{listings}

\begin{document}

%%
%% The "title" command has an optional parameter,
%% allowing the author to define a "short title" to be used in page headers.
\title{Why Does My Transaction Fail? A First Look at Failed Transactions on the Solana Blockchain}

\author{Xiaoye Zheng}
\orcid{0009-0001-9048-1930}
\affiliation{%
  \institution{Zhejiang University}
  \department{The State Key Laboratory of Blockchain and Data Security}
  \city{Hangzhou}
  \country{China}
}
\email{xiaoyez@zju.edu.cn}

\author{Zhiyuan Wan}
\orcid{0000-0001-7657-6653}
\affiliation{%
  \institution{Zhejiang University}
  \department{The State Key Laboratory of Blockchain and Data Security}
  \city{Hangzhou}
  \country{China}
}
\email{wanzhiyuan@zju.edu.cn}
\authornote{Zhiyuan Wan is the corresponding author.}
\authornote{Also with Hangzhou High-Tech Zone (Binjiang) Institute of Blockchain and Data Security.}

\author{David Lo}
\orcid{0000-0002-4367-7201}
\affiliation{%
  \institution{Singapore Management University}
  \department{School of Computing and Information Systems}
  \city{Singapore}
  \country{Singapore}
}
\email{davidlo@smu.edu.sg}

\author{Difan Xie}
\orcid{0009-0005-5777-1679}
\affiliation{%
  \institution{Hangzhou High-Tech Zone (Binjiang) Institute of Blockchain and Data Security}
  \city{Hangzhou}
  \country{China}
}
\email{xiedifan@bcds.org.cn}

\author{Xiaohu Yang}
\orcid{0000-0003-4111-4189}
\affiliation{%
  \institution{Zhejiang University}
  \department{The State Key Laboratory of Blockchain and Data Security}
  \city{Hangzhou}
  \country{China}
}
\email{yangxh@zju.edu.cn}

\titlenote{This research was supported by the National Science Foundation of China (No. 62472383 and No. 62102358), and the Open Research Fund of the State Key Laboratory of Blockchain and Data Security, Zhejiang University.}

\begin{abstract}
Solana is an emerging blockchain platform, recognized for its high throughput and low transaction costs, positioning it as a preferred infrastructure for Decentralized Finance (DeFi), Non-Fungible Tokens (NFTs), and other Web 3.0 applications. 
In the Solana ecosystem, transaction initiators submit various instructions to interact with a diverse range of Solana smart contracts, among which are decentralized exchanges (DEXs) that utilize automated market makers (AMMs), allowing users to trade cryptocurrencies directly on the blockchain without the need for intermediaries.
Despite the high throughput and low transaction costs of Solana, the advantages have exposed Solana to bot spamming for financial exploitation, resulting in the prevalence of failed transactions and network congestion.

Prior work on Solana has mainly focused on the evaluation of the performance of the Solana blockchain, particularly scalability and transaction throughput, as well as on the improvement of smart contract security, leaving a gap in understanding the characteristics and implications of failed transactions on Solana. 
To address this gap, we conducted a large-scale empirical study of failed transactions on Solana, using a curated dataset of over 1.5 billion failed transactions across more than 72 million blocks. 
Specifically, we first characterized the failed transactions in terms of their initiators, failure-triggering programs, and temporal patterns, and compared their block positions and transaction costs with those of successful transactions.
We then categorized the failed transactions by the error messages in their error logs, and investigated how specific programs and transaction initiators are associated with these errors.

We find that transaction failure rates on Solana exhibit recurring daily patterns, and demonstrate a strong positive correlation with the volume of failed transactions, with bots on Solana experiencing a high transaction failure rate of 58.43\%. We identify ten distinct error types in the error logs of failed transactions, with \textit{price or profit not met} and \textit{invalid status} errors accounting for 67.18\% of all failed transactions. AMMs primarily experience \textit{invalid status} errors among failed transactions, while DEX aggregators are more commonly affected by \textit{price or profit not met} errors. Among transaction initiators, bots encounter a broader range of errors due to their high-frequency trading and complex interactions with smart contracts. In contrast, human users experience a more limited range of errors.
Based on our findings, we provide recommendations to mitigate transaction failures on Solana and outline future research directions.
\end{abstract}




\maketitle

\section{Introduction}\label{sec:introduction}

Solana is an emerging blockchain platform recognized for the high throughput and low transaction costs that it offers, making it a preferred infrastructure for DeFi~\cite{chen2020blockchain}, NFTs~\cite{chohan2021non}, and other applications of Web 3.0~\cite{hendler2009web3}.  
At the core of the design of Solana is a stateless smart contract model, where smart contracts (i.e., Solana programs) do not retain internal states. Instead, all necessary state data resides in external accounts, which are passed as inputs when transactions invoke these programs. This stateless model not only streamlines the logic of programs and minimizes storage on-chain, but also facilitates the parallel execution of transactions. Together with the Proof of History mechanism and the Tower BFT consensus protocol of Solana~\cite{solana_consensus}, the design achieves a substantial increase in throughput while maintaining low transaction fees.

\begin{figure}[t]
    \centering
    \begin{minipage}[b]{0.48\linewidth}
        \centering
        \includegraphics[width=\linewidth]{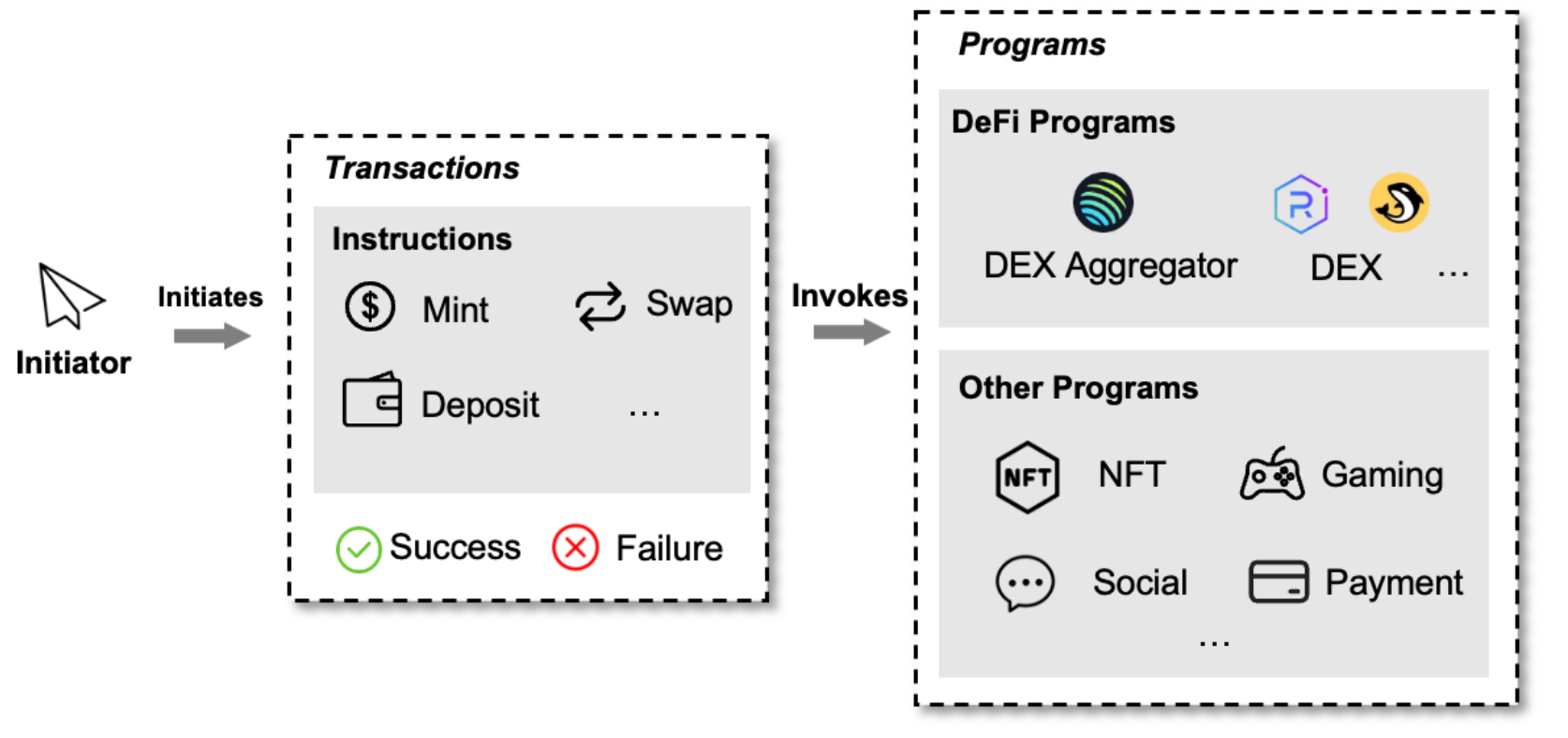}
        \caption{Overview of Solana Ecosystem.}
        \label{fig:bg}
    \end{minipage}
    \hfill
    \begin{minipage}[b]{0.48\linewidth}
        \centering
        \includegraphics[width=0.9\linewidth]{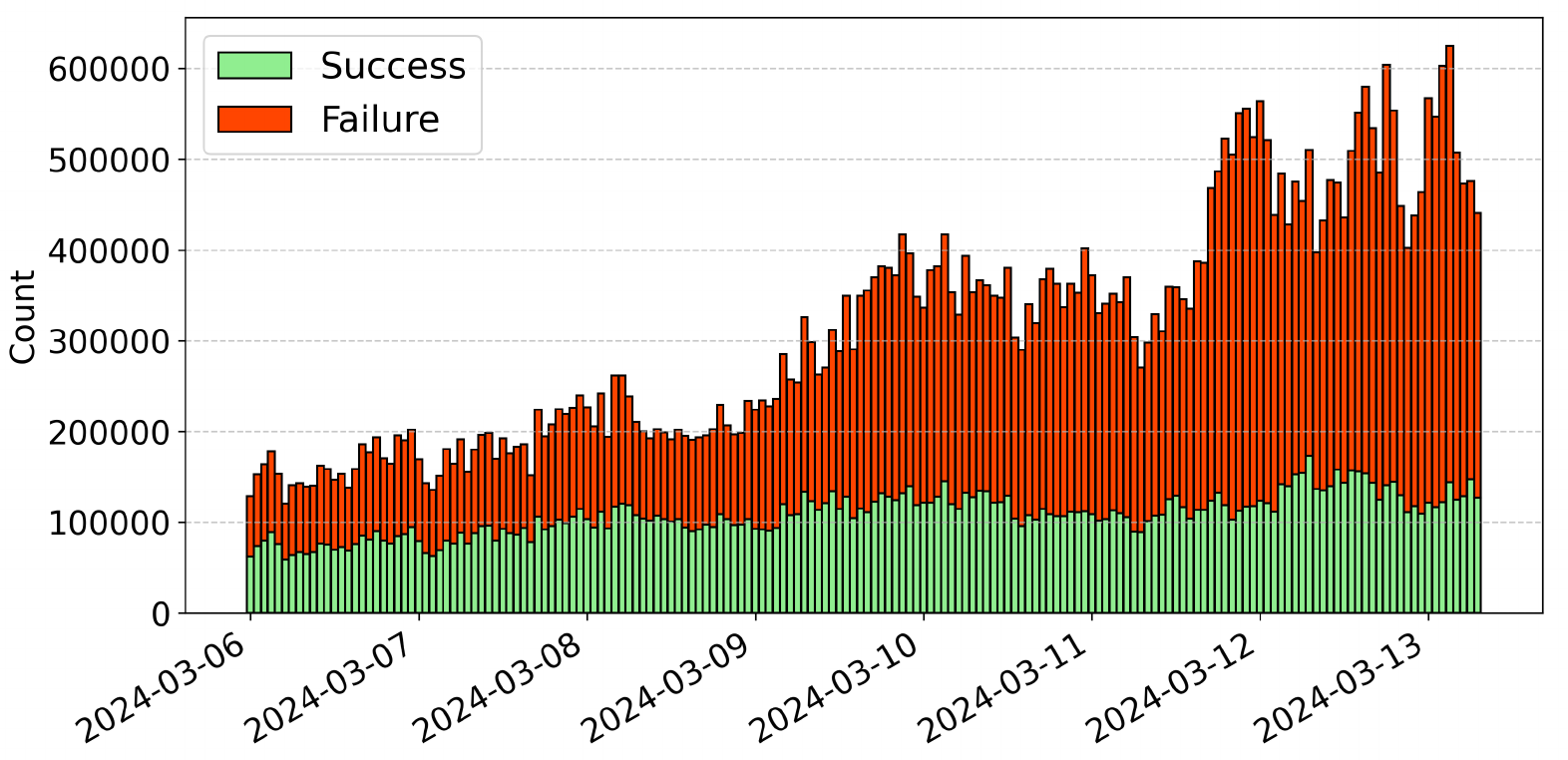}
        \caption{Hourly Trends of Successful and Failed Non-Vote Transactions on Solana (March 6-13, 2024).}
        \label{fig:failed_cnt}
    \end{minipage}
    \vspace{-10pt}
\end{figure}

Figure~\ref{fig:bg} presents a high-level overview of the Solana ecosystem, illustrating the typical interaction flow from transaction initiators to the various Solana programs they interact with. On the left, transaction initiators submit various instructions, such as minting, swapping, and depositing. A transaction is considered \textit{successful} if all instructions execute without errors; otherwise, it is classified as \textit{failed} if one or more instructions encounter errors during execution.
On the right, the figure illustrates the range of programs invoked by Solana transactions, including DeFi programs such as DEXs and DEX aggregators. DEXs enable users to trade assets directly with one another on a blockchain, without the need for centralized intermediaries. DEXs frequently rely on AMM mechanisms, which allow users to trade assets through liquidity pools rather than traditional order books.

While Solana aims to maintain low transaction fees and high throughput, the low cost of transactions has inadvertently exposed the Solana network to bot spamming for financial exploitation~\cite{bots1, bots2}, where transactions are automatically generated by machines. The bot activity has led to increased failure rates of transactions and network congestion, with reports indicating that up to 75\% of non-vote transactions failed during periods of ``memecoin mania''~\cite{memecoin} and bot-induced spam. Non-vote transactions, unlike vote transactions required for network consensus, encompass general-purpose activities such as token transfers and smart contract executions, making them more vulnerable to automated abuse. Mert Mumtaz, CEO of Helius, notes that a significant portion of these transaction failures originates from bot activity rather than normal user interactions, resulting in a degraded user experience on the network~\cite{failed_tx_solana}.
Our experimental results, depicted in Fig.~\ref{fig:failed_cnt}, present the hourly trends of failed non-vote transactions on Solana from March 6 to March 21, 2024, highlighting the prevalence of transaction failures on Solana.

Previous studies on the Solana ecosystem have primarily explored two aspects of the ecosystem, either evaluating the performance of the Solana blockchain, with a particular focus on its scalability~\cite{pierro2022scalability} and transaction throughput~\cite{duffy2021IoT, pierro2022can}, or enhancing smart contract security through vulnerability detection techniques~\cite{vrust, smolka2023fuzz}. Nonetheless, these efforts have overlooked the transaction-level analysis in the Solana ecosystem, leaving a critical gap in understanding the characteristics and implications of failed transactions on Solana.
As reported in previous studies~\cite{cryptoNewsFailedAnalysis,duneFailedAnalysis}, 
Solana exhibits a significantly higher transaction failure rate, which can exceed 75\% during periods of network congestion, in contrast to the transaction failure rate 
from 1\% to 3\% on Ethereum
~\cite{oliveira2021analyzing,eth-failed-rate}.
Such high transaction failure rate of Solana can introduce delays in transaction finality and contribute to an increased network load, potentially disrupting the seamless operation expected by users and developers. If left unaddressed, high transaction failure rates could undermine confidence of users and developers in the Solana ecosystem, particularly if users and developers come to perceive the ecosystem as unreliable. Thus, 
a comprehensive understanding of the characteristics and implications of failed transactions would provide valuable insights into the Solana ecosystem, enabling both regular users and bot developers to optimize transaction design and execution, ultimately improving the efficiency of the transactions and the reliability of the network.

To address this gap, we conducted a large-scale empirical investigation into failed transactions on the Solana blockchain. Specifically, we curated a dataset comprising 2,898,175,006 non-vote transactions, comprised of 1,387,340,838 \emph{successful} and 1,510,834,168 \emph{failed} transactions, spanning from August 1, 2023 to July 31, 2024, enriched with both relevant on-chain and off-chain data. Below, we present our research questions and highlight the key findings:

\noindent\textbf{RQ1: What are the characteristics of the failed transactions?} 

Given the high transaction volume and substantial failure rates on Solana, it is essential to systematically capture the characteristics of these failed transactions. The understanding of the failed transactions can help identify issues that contribute to the failures, and inform the development of targeted strategies to address these issues in the Solana ecosystem. Therefore,  we conducted a two-level analysis of the failed transactions on the Solana blockchain in our dataset, macro- and micro-level analysis. At the macro level, we captured the initiators, failure-triggering programs, and temporal patterns of the failed transactions. 
At the micro level, we focused on attributes of the transactions in our dataset, comparing transaction fees and block positions between successful and failed transactions. 
Our analysis revealed distinct daily patterns in transaction failures, with bots showing a high transaction failure rate of 58.43\%. The top ten programs with the highest volume of failed transactions account for 77.95\% of all failed transactions. In addition, failed transactions tend to be positioned deeper in blocks\footnote{A Solana block is a collection of transactions processed and confirmed by validators on the Solana blockchain, representing a specific, sequential segment of the transaction history of the blockchain.} as compared to successful transactions (592 vs. 529 in median). 

\noindent\textbf{RQ2: What are the errors that cause transaction failures?} 

Failed transactions on Solana are accompanied by a range of errors, as outlined in their log messages. These errors can provide insights into the underlying causes of transaction failures, and uncover programming challenges faced by developers and users of Solana. Thus, in this RQ, we aim to establish a systematic classification of the errors responsible for transaction failures in our dataset. Specifically, we extracted error messages from the log messages associated with failed transactions in our dataset, and applied thematic analysis to categorize the identified error messages into distinct error types.
We identified 10 distinct error categories, with the vast majority of failed transactions (67.18\%) attributed to either \textit{price or profit not met} or \textit{invalid status} errors. A smaller portion of failures stems from issues arising in the improper interactions between transaction initiators and the corresponding programs, including \textit{out of funds} (2.16\%), \textit{invalid input parameters} (2.55\%), and \textit{invalid input account} (3.27\%). Furthermore, a limited number of failures are linked to runtime environment issues (0.72\%) and resource management constraints, with \textit{out of resource} errors accounting for 0.49\%, indicating potential gaps in program testing and validation for initiating Solana transactions.

\noindent\textbf{RQ3: How do programs and transaction initiators contribute to different error types in transaction failures?}

Transactions on Solana involve interactions between programs and transaction initiators, with failures arising from both program design and the actions of transaction initiators. Characterizing the programs and transaction initiators contribute to transaction failures can provide insights into improving program design and transaction initiation practices, with the goal of reducing failures by addressing specific error types. Thus,
we investigated how programs and transaction initiators shape the distribution of error types in failed transactions. Our investigation focused on two dimensions, the error type distributions across the programs with high failed transaction volumes, as well as the difference in distributions of error types between bot-initiated and human-initiated transactions. 
We observed that the programs of DEX aggregators predominantly fail due to \textit{price or profit not met} errors. The programs of AMM frequently encounter \textit{invalid status} errors, often caused by bots attempting to front-run or manipulate transactions. 
Bot-initiated transactions exhibit various error types, with \textit{price or profit not met} errors being the most common.  Human-initiated transactions are more prone to \textit{out of funds} errors as compared to bot-initiated transactions.

Based on our findings, we discuss implications and provide recommendations for Solana ecosystem core developers, protocol designers, and end users, including mechanisms and tools to mitigate transaction failures, and outline future avenues of research. We make the following contributions:
\begin{itemize}[nosep]
    \item We present the first large-scale empirical study on failed transactions on Solana, identifying 10 distinct error types associated with failed transactions and providing insights into the Solana ecosystem.
    \item We curate a dataset that comprises 1,510,834,168 failed transactions on Solana along with corresponding error messages for future investigation by others.
    \item We provide practical recommendations for practitioners in the Solana ecosystem, and outline future avenues of research.
\end{itemize}

\section{Background}\label{sec:background}
The architecture of the Solana blockchain is built around four essential concepts. To interact with the Solana blockchain, an entity must first create an \textit{account}, which serves not only as a unique identifier but also as a store of relevant data and states. This account can initiate the execution of a smart contract, referred to as a Solana \textit{program}, by submitting a \textit{transaction}. A transaction consists of one or more \textit{instructions}, each specifying an operation to be performed by a program. The following provides a detailed exploration of these concepts.

\subsection{Solana Accounts and Programs}

The Solana blockchain relies on a flexible account-based model to manage both data and executable logic. At the core of the model are two distinct types of accounts: (1) \emph{data accounts}, which store non-executable data, and (2) \emph{program accounts}, which hold executable code that defines the behavior of programs. Programs in Solana are stateless, they do not maintain persistent data between transactions. Instead, they interact with data accounts to store and modify state(s) as needed.

Each account in Solana is uniquely identified by a 32-byte public key and consists of four components~\cite{solana_doc}:
(1) Public key of the owner program, which specifies the program that controls the operations of the account.
(2) Executable flag, which indicates whether the account contains a program (executable code).
(3) Data content, which stores relevant information, such as token balances or configuration parameters.
and (4) Balance, which holds the balance of the account, measured in Solana's smallest currency unit, lamports.

In the stateless programming model of Solana, programs manage states by creating or updating data accounts rather than maintaining states internally. 
Consequently, for each instruction of a transaction to execute, the initiator of the transaction must explicitly provide all relevant accounts as instruction arguments, including both the data accounts and program accounts involved in the operation.

\subsection{Solana Transactions}
In Solana, transactions are divided into two categories, \emph{vote} transactions and \emph{non-vote} transactions, each serving a distinct role in the Solana network. 
Vote transactions are system-level operations initiated by validator nodes to maintain network consensus by confirming new blocks and ensuring synchronization of the blockchain's state. Vote transactions are essential for preserving network stability and are integral to the consensus mechanism of Solana~\cite{vote_tx}.
In contrast, non-vote transactions encompass all user-initiated activities, such as token transfers, account creation, decentralized exchange (DEX) operations, and NFT minting. Non-vote transactions represent the primary interactions of users and applications with Solana~\cite{non_vote_tx}.
The separation between vote and non-vote transactions aims to ensure that the network can efficiently handle both consensus-related operations and user-driven activities, maintaining high throughput and reliability of Solana.

Solana transactions are composed of \emph{instructions} that interact with various on-chain programs, where each instruction specifies a distinct operation to be performed~\cite{instruction}.
When an entity submits a transaction to the Solana network, it includes one or more instructions, each containing the program to be invoked, relevant account addresses, and a data byte array. The designated program processes the data and executes the specified operations on the given accounts. Transactions on Solana are executed sequentially and atomically--if any instruction fails, all state changes made by the transaction are discarded, ensuring system consistency~\cite{squential_execution}.

The lifecycle of a transaction begins when it is \textbf{\textit{submitted}} to a Remote Procedure Call (RPC) server--an endpoint that allows entities to interact with the blockchain network--which forwards it to both the current leader and the next scheduled leader. 
Note that a leader is a validator temporarily assigned to produce blocks during a specific time slot. Unlike Ethereum, where transactions are propagated through a public mempool, Solana broadcasts transactions directly to validators responsible for block production. The leader processes the transaction, appends it to the current block, and then propagates it across the network using Solana's Turbine broadcast mechanism, which efficiently fragments and distributes data to validators~\cite{block_propagation}.
For a transaction to be \textbf{\textit{accepted}} and \textbf{\textit{processed}}, it must contain one or more \textbf{\textit{signers}} who authorize the operation by providing digital signatures generated with their private keys.
Each transaction must include at least one writable signer account, typically the fee payer, who bears the transaction cost. The signer account is serialized first in the account input list of the transaction to ensure the fee deduction happens before any further processing. The signer accounts can be categorized into two distinct types~\cite{niedermayer2024bots}: (1) bot accounts (\faRobot), which are operated programmatically by automated systems, and  
(2) human accounts (\faUser), which are managed manually by users.

A block is considered \textbf{\textit{confirmed}} once it receives votes from at least two-thirds of the active validators. For a transaction to reach \textbf{\textit{finality}}, 31 additional blocks must be built upon the corresponding block containing the transaction, ensuring its immutability~\cite{tx_finality}.
A finalized transaction can be classified as either \textbf{\emph{successful}} if all its instructions are executed without any errors, or \textbf{\emph{failed}} if one or more instructions in it encounter an error during its execution.
Failed transactions produce detailed log messages, enabling developers to diagnose the underlying issues. For instance, Listing~\ref{lst:log}
presents a log message fragment from the failed transaction \texttt{36DLJ5vufG}, in which, the error log entry at line 6 indicates a slippage-related error. We attribute responsibility for the error to the outermost program in the call stack of the log message, which in this example is the program \texttt{6Q4Xu2}, as identified at line 1 of Listing~\ref{lst:log}.

Both successful and failed transactions incur fees to cover the operational costs of processing and prevent spam in the Solana network. Each transaction requires a base fee of 5,000 lamports per signature, regardless of the computational resources required, ensuring that even small transactions contribute to the sustainability of the network~\cite{solana_doc_fee}. 
Lamports are the smallest denomination of Solana's native cryptocurrency, SOL, where 1 SOL equals 1 billion lamports. 
Solana provides further flexibility through two parameters, \textit{compute unit limit} and \textit{compute unit prices}, allowing users to optimize the performance and priority of the transactions based on their complexity and urgency.
The compute unit limit specifies the maximum computational resources, measured in compute units (CU), that a transaction can consume. The limit is configured by the sender of a transaction based on the complexity of the transaction. The compute unit price, measured in micro-lamports per CU, reflects the extra amount of SOL that the sender is willing to pay for a transaction. Validators are incentivized to process transactions with higher CU prices first, ensuring optimal resource utilization in the network and faster execution of transactions. 

Solana employs a fee-based prioritization system~\cite{priority_fees} to optimize transaction processing to maintain the high performance of Solana. While users can increase transaction fees to compete for higher priority in transaction processing, the transaction scheduling mechanism of Solana does not always guarantee that transactions with higher fees be processed in prior to those with lower fees~\cite{solana_fees_in_practice}. Other factors, such as the allocation of compute units by transaction initiators~\cite{solana_doc_cu} and the mechanisms applied by validators~\cite{solana_fees_in_practice}, can affect the order of transactions in blocks on Solana. For instance, the Jito Tips mechanism allows searchers and users to offer tips to validators in return for prioritizing their transactions~\cite{solana_tx_latency}.

\begin{lstlisting}[caption={Log Message Fragment of an Example Failed Transaction on the Solana Blockchain.},label=lst:log,xleftmargin=20pt, xrightmargin=20pt]
Program 6Q4Xu2 invoke [1]
Program log: Instruction: Swap
Program log: x.com/TechnoBotSolana
Program 675kPX9 invoke [2]
Program log: ray_log: ...
Program log: Error:exceeds desired slippage limit
Program 675kPX9 consumed 16323 of 62684 compute units
Program 675kPX9 failed: custom program error: 0x1e
Program 6Q4Xu2 consumed 53339 of 99700 compute units
Program 6Q4Xu2 failed: custom program error: 0x1e
\end{lstlisting}

\section{Dataset}\label{sec:methodology}
\subsection{Data Collection and Preprocessing}
\label{sec:data_collect}

To conduct a comprehensive investigation of failed transactions on Solana, we selected a one-year period from August 1, 2023 to July 31, 2024, spanning 72,123,900 blocks on the Solana blockchain with the block range from 208,703,000 to 280,826,900. Given the large transaction volume over the one-year period, we designed a stratified sampling strategy for block selection on a weekly basis, where we randomly selected one day per week throughout the year, while ensuring a balanced representation of both weekdays and weekends. The sampling strategy achieves a 99.999\% confidence level with a margin of error of ±0.06\%, ensuring that the selected sample represents the broader transaction patterns on Solana. 
Consequently, we selected 10,458,452 blocks from 53 days between August 1, 2023 and July 31, 2024.

\noindent\textbf{On-chain Data.}
We first retrieved the selected blocks from the Solana blockchain through an RPC node, provided by Alchemy~\cite{alchemy}, utilizing the Solana JSON RPC API.
For each block, we iterated over its transactions, filtered out vote transactions (those invoking the vote program) and extracted multiple entries from each remaining non-vote transaction, including (1) the block number, (2) the block timestamp, (3) transaction hash, (4) account inputs, (5) signer(s), (6) transaction fees, (7) \emph{compute units} the transaction costs, (8) transaction log messages, and (9) program address and its corresponding instruction that raises errors if the transaction fails. Note that we consider a transaction as a \emph{failed} transaction if the error fields in its metadata are not empty. For each failed transaction, we further analyzed the log messages to identify the specific error log messages the triggered during execution.

\noindent\textbf{Off-chain data of Solana.} 
We also collected off-chain data related to Solana programs. We obtained program public names from Solscan~\cite{solscan}, a comprehensive Solana blockchain explorer and analytics platform. Additionally, we retrieved and cloned program source code from GitHub repositories when available through URLs listed in Solscan's security section.

Consequently, we curated a dataset of 2,898,175,006 non-vote transactions on the Solana blockchain, comprising both on-chain data (including block information, transaction details, compute units, and error messages if any) and off-chain data (including program names and their source code if any).

\subsection{Account Classification}
The transactions on Solana may originate from accounts managed by machines, bot accounts, or human users.
To classify the accounts that initiate the transactions in our dataset, we applied a classification algorithm that takes into account the on-chain behavior of the accounts. Specifically, we characterized the on-chain behavior of a Solana account by considering multiple features from its transaction histories, which aligned with prior work~\cite{huang2020eosio, niedermayer2024bots}:

\noindent \blackcircle{1} \textbf{\emph{Transaction Frequency:}} We measured the transaction frequency of each account by computing the average and variance of the intervals between the transactions the account has initiated. Specifically, for each account, we extracted a time series $[t_1, t_2, ..., t_n]$, representing the timestamps of $n$ transactions initiated by the account within our dataset. $t_i$ denotes the timestamp when the account initiated its $i$th transaction, as recorded in the corresponding block. The interval between consecutive transactions is calculated as $t_{i+1}-t_{i}$, where $i < n$. 

\noindent \blackcircle{2} \textbf{\emph{Transaction Volume: }}
We quantified the transaction volume of each account by calculating both the total number of transactions and the average number of transactions per block initiated by the account within our dataset. In calculating the average transactions per block, we excluded blocks that do not contain any transactions initiated by the respective account. 

We utilized the scikit-learn implementation\footnote{\url{https://scikit-learn.org/}} of the random forest (RF) algorithm~\cite{pal2005random} to classify 12,712,516 accounts in our dataset. Specifically, we first constructed the ground truth by manually labeling 200 accounts randomly selected from our dataset. Specifically, for each account, we analyzed the frequency, volumes and intervals of transactions the account initiated, programs the account interacted with, and its involvement in casino games. The manual labeling resulted in the ground truth consisting of 96 human accounts and 104 bot accounts. Using the ground truth as the training set, we built a RF classification model to categorize the remaining accounts. For each account, the classification model outputs a probability for the human or bot category. Accounts with a probability above 0.9 were classified with high confidence, while those below 0.9 were labeled as ``unknown''. As a result, we classified the accounts in our dataset into 803,136 bot accounts and 1,359,772 human accounts. 
To evaluate the reliability of the classification model, we randomly sampled 97 bot accounts and 97 human accounts from the classifications for validation, using a 95\% confidence interval and a 10\% sampling error. Among the 194 randomly sampled accounts, we identified six misclassified cases, including three human accounts exhibiting high-frequency trading within a single day, and three bot accounts performing low-frequency but consistent transactions, suggesting that our classification model achieved an accuracy of 96.91\%.

\vspace{-5pt}
\section{Characteristics of Failed Transactions (RQ1)}\label{sec:RQ1}
\subsection{Methodology}

To answer RQ1, we first aggregated the failed transactions based on the accounts that initiated the transactions, distinguishing between bots and human users. 
Next, we grouped the failed transactions by the programs responsible for the corresponding failures and characterized the functionality of these programs. 
To account for variations in overall transaction volume, we also normalized the number of failed transactions against the total non-vote transactions per hour, yielding the hourly transaction failure rate.
Lastly, we explored the relationship between hourly transaction volume and failure rate, and employed autocorrelation analysis~\cite{autocorrelation} to uncover potential periodic patterns in the time series of transaction failure rates.

We followed up with a micro-level analysis of the failed transactions in our dataset. 
Specifically, we first compare the positions within blocks, as well as the transaction fees, compute units, and cost efficiency between successful and failed transactions on Solana. We then employed Wilcoxon rank-sum tests to determine the statistical significance of these differences.

\vspace{-5pt}
\subsection{Results}
\subsubsection{Macro-Level Analysis} We first provide an overview of the failed transactions on Solana from three dimensions: the types of accounts initiating these transactions, the programs responsible for the failures, and the temporal trends in transaction volume and failure rate.

\noindent\textbf{Account Types.}
Table~\ref{tab:signer} compares identified bot and human accounts on the Solana blockchain, revealing notable differences in the numbers of accounts and failure rates of transactions between bot and human accounts. Bot accounts, comprising over 0.8 million accounts in our dataset, exhibit a high transaction failure rate of 58.43\%, with approximately 453.5 million failed transactions compared to 322.6 million successful ones. The high failure rate suggests that bot-initiated transactions may rely heavily on high-frequency strategies, which may contribute to network congestion. 
In contrast, the 1,359,772 human accounts show a considerably lower transaction failure rate of 6.22\%, achieving 2,978,711 successful transactions against only 197,535 failures. 
The disparity in transaction failure rates between bot and human accounts indicates that human-initiated transactions are more deliberate, potentially benefiting from strategic resource management and reduced exposure to network contention, as compared to bot-initiated transactions.
\begin{table}
    \centering
\caption{Comparison of Failed and Successful Transactions Initiated by Bot and Human Accounts on Solana.}
\vspace{-10pt}
    \resizebox{0.9\linewidth}{!}{
    \begin{tabular}{cccc}

    \toprule
        \textbf{Account Type} & \textbf{\# Identified Accounts} & 
        \textbf{\# Failed Transactions} & \textbf{\# Successful Transactions} \\
        % \multicolumn{2}{c}{\textbf{Number of Transactions (Txs)}} \\
         % &  & \emph{\# Failed Txs}  &\emph{\# Successful Txs}\\
        \midrule
         Bot      & 803,136  & 453,547,307 & 322,618,527\\
         Human    & 1,359,772    & 197,535 &   2,978,711\\
    \bottomrule

    \end{tabular}
    \vspace{-20pt}
    }
    \label{tab:signer}

\end{table}

\rqbox{
\textbf{Finding 1:} 
Bot accounts on Solana exhibit a high transaction failure rate of 58.43\%, in contrast to the substantially lower rate of 6.22\% observed among human accounts.  The disparity indicates that human-initiated transactions tend to be more deliberate than automated high-frequency bot operations.
}

\begin{figure*}[h!]
    \centering
    \vspace{-15pt}
    \begin{minipage}{0.28\textwidth}
        \centering
        \includegraphics[width=\linewidth]{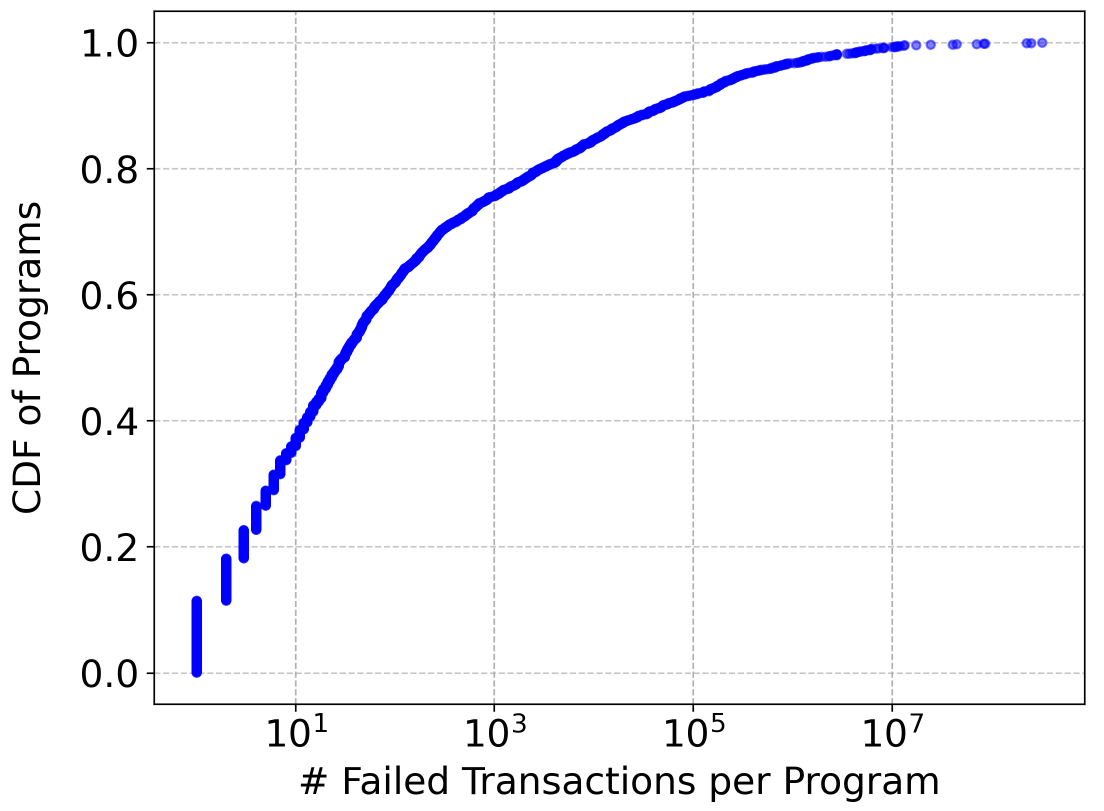}
         \vspace{-20pt}
        \caption{Cumulative Distribution of Programs Responsible for Failed Transactions on Solana.}
        \label{fig:failed_program_cdf}
    \end{minipage}%
    \hfill
    \begin{minipage}{0.6\textwidth}
        \centering
    \includegraphics[width=\linewidth]{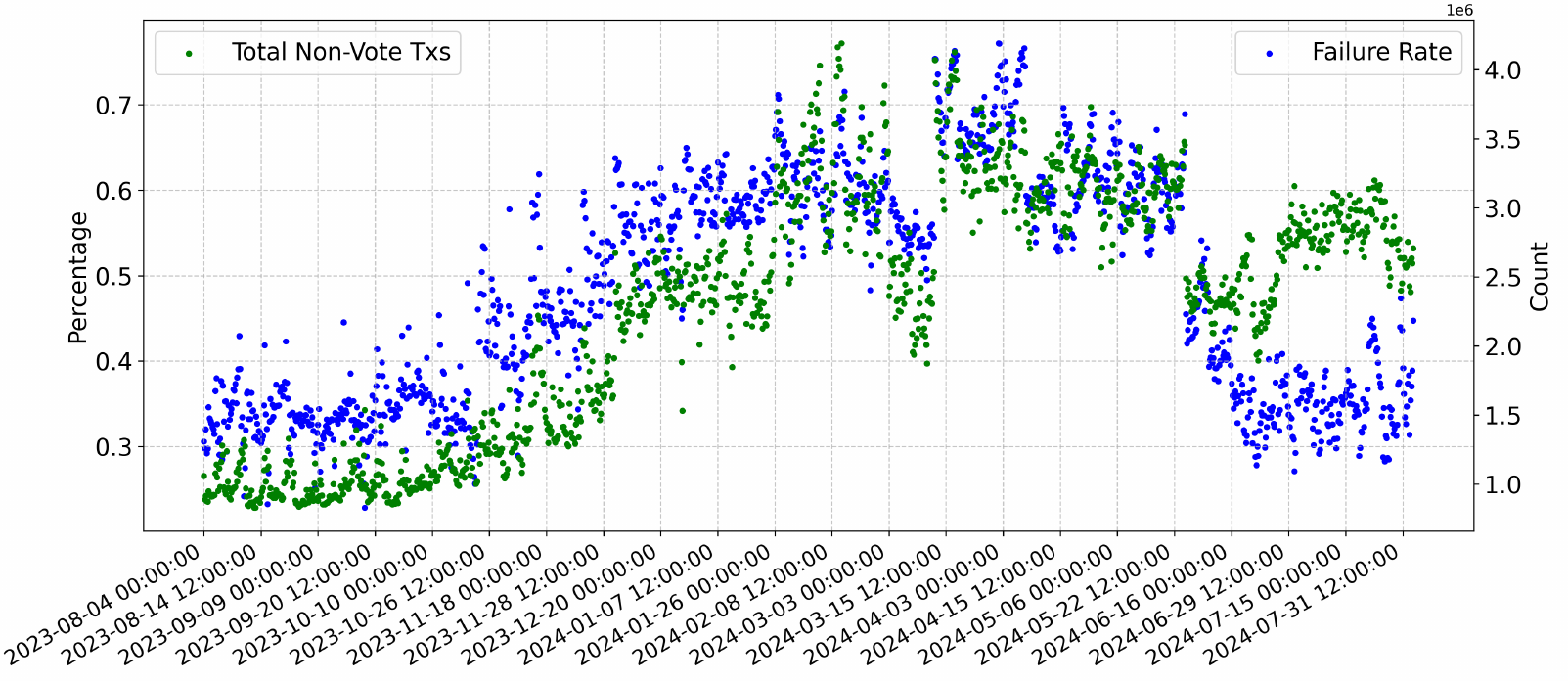}
     \vspace{-20pt}
\caption{Hourly Number of Non-Vote Transactions and Failure Rate on Solana.}
    \label{fig:failed_ratio}
    \end{minipage}
    \vspace{-10pt} 
\end{figure*}

\noindent\textbf{Programs.} A total of 2,324 distinct programs are identified as being responsible for at least one failed transaction in our dataset, among which, each program is responsible for a median of 30 failed transactions (min: 1, average: 649,411, max: 327,650,466).
Figure~\ref{fig:failed_program_cdf} presents the logarithmic scale Cumulative Distribution Function (CDF) of the number of failed transactions across these programs. 
The CDF curve exhibits a steep increase from $10^1$ to approximately $10^3$ failed transactions, indicating that a large number of programs are responsible for relatively few failed transactions. 
After the initial steep increase, the slope of the curve becomes more gradual, indicating that fewer programs are responsible for higher numbers of failed transactions.
The curve extends towards the rightmost end, suggesting that a very small fraction of programs are responsible for hundreds of millions of failed transactions, which tend to play critical roles in the Solana ecosystem.

\begin{table*}[]
    \centering
\caption{Top 10 Programs Responsible for the Highest Volume of Failed Transactions (Txs) on Solana.}
\vspace{-10pt}
    \resizebox{0.9\linewidth}{!}{        
    \begin{tabular}{cllcp{1.5cm}p{2cm}}
    \toprule

        \textbf{Address} & \textbf{Public Name} & \textbf{Description} &\textbf{Failed Txs (\# | \%)}  & \textbf{Failure Rate (\%)} & \textbf{Affected Accounts (\#)}\\
        \midrule
         675kPX9MHT   & Raydium Liquidity Pool V4~\cite{raydium} & AMM         & 327,650,466 | 21.69\%   & 74.21\% & 1,565,324 \\
         JUP6LkbZbj    & Jupiter Aggregator V6~\cite{JupiterV6}  & DEX  aggregator             & 252,017,540 | 16.68\% &  79.74\% &1,115,631\\
         cjg3oHmg9uu   & Chainlink Data Store~\cite{chainlink}                    & Decentralized Data Storage         & 226,952,883 | 15.02\% & 94.18\% & 23\\
         6Q4Xu2sXxM   & -                    & -         & 87,172,461 | 5.77\% & 99.12\% &105 \\
         JUP4Fb2cqi    & Jupiter Aggregator V4~\cite{JupiterV4}      &DEX  aggregator         & 84,279,822 | 5.58\%  & 96.14\% &5,816 \\
         GDDMwNyyx8   & Sequence Enforcer~\cite{sequence_enforcer}         & Transaction ordering mechanism     & 71,616,610 | 4.74\% & 37.37\% &105\\
         
         ATokenGPvb   & Associated Token Account~\cite{ATA} & Account management utility   & 44,956,578 | 2.98\%& 32.90\% &544,818 \\
         
         9uW2TqLyfY   & -          & -               & 40,817,555 | 2.70\% & 97.67\% &924\\

         YmirFH6wUr   & -          & -               & 24,635,690 | 1.63\% & 95.21\% &4\\
    
         3J3HFc8jXx   & -              & -               & 17,509,217 | 1.16\%& 91.75\% & 7 \\
         
    \bottomrule
    \end{tabular}
    }
    \label{tab:programs}
   \vspace{-15pt}
\end{table*}
We further performed a detailed analysis of the top 10 programs responsible for the highest volume of failed transactions in our dataset, situated at the rightmost end of the CDF curve. An overview of the ten programs is presented in~Table~\ref{tab:programs}, among which, six are public services, while four are privately deployed with no name disclosed. We made the following observations:
\begin{itemize}[nosep]
    \item Operating as an automated market maker (AMM)\footnote{An automated market maker is a decentralized trading mechanism that allows users to trade assets through liquidity pools rather than traditional order books. In an AMM, liquidity providers deposit pairs of tokens into these pools, and trades are facilitated by an algorithm that adjusts prices based on the pool's token ratios. This model is widely used in DEXs, providing continuous liquidity and enabling permissionless trading.}, Raydium Liquidity Pool V4 alone accounts for 21.69\% of all failed transactions, totaling over 
    327 million failed attempts, affecting 1,565,324 accounts. The high failure rate of 74.21\% indicates that a vast number of transactions interacting with the program encounter issues, potentially due to liquidity constraints, programmatic bugs, or excessive competition for resources.

    \item Jupiter Aggregator V6 follows closely, designed as a decentralized exchange (DEX) aggregator, contributing 16.68\% of the total failed transactions with a failure rate of 79.74\%, affecting 1,114,631 accounts. %(61\%), the highest number of affected accounts among all programs listed.

    \item The Chainlink Data Store program contributes to 15.02\% of all failed transactions, with the high failure rate of 94.18\% being attributed to only 23 accounts.

    \item The Associated Token Account and Sequence Enforcer programs, both integral to the blockchain operations of Solana, exhibit a comparable moderate failure rate, each approaching 40\%. This indicates potential bottlenecks at the infrastructure level of Solana, such as network congestion or insufficient computational resources. 
    Meanwhile, a significant difference exists in the number of accounts they affect (544,818 vs. 105), reflecting the fundamental difference in their usage patterns and operational scope in the Solana ecosystem. In particular, the Associated Token Account program facilitates everyday user interactions and token management, while the Sequence Enforcer program focuses on transaction order enforcement in specialized contexts.

    \item Several programs, such as the one associated with the address 6Q4Xu2sXxM and Jupiter Aggregator V4, display extremely high failure rates, exceeding 95\%. This suggests they might be highly congested or not adequately optimized for the volume of incoming transactions.
\end{itemize}

\rqbox{
\textbf{Finding 2:} 
50\% of the programs were involved in fewer than 30 failed transactions on Solana. The top ten programs with the highest volume of failed transactions account for 77.95\% of all failed transactions, with three of them being public DeFi programs in the Solana ecosystem.}

\noindent\textbf{Temporal Trends.}
Figure~\ref{fig:failed_ratio} provides a detailed temporal analysis of non-vote transactions on the Solana blockchain, depicting the sampled transaction volume (green dots) and transaction failure rate (blue dots) on an hourly basis between August 1, 2023 and July 31, 2024. Overall, the fluctuations in non-vote transaction volume appear to coincide with the variations in transaction failure rates over time.
% %
On the one hand, the hourly transaction volume (green dots) demonstrates noticeable variability over time, with frequent peaks and troughs. The variability may suggest periodic surges in network activity, possibly driven by specific applications, market events, or high-demand use cases like token transfers or DeFi operations.
% %
On the other hand, the hourly transaction failure rate (blue dots) similarly fluctuates, indicating that higher failure rates coincide with increased transaction volume.
% %
It is noteworthy that the transaction failure rates experienced a significant reduction after June 16, 2024, accompanied by a modest decrease in transaction volume. The reduction in transaction failure rates may be attributed to the validator update and consensus optimization that were introduced to the Solana blockchain on June 10, 2024~\cite{twitter_solana_status}.
A Wilcoxon rank-sum test confirms the statistical significance of the difference in block positions between failed and successful transactions  (p-value < 0.001), suggesting that failed transactions are outcompeted, leading to their deeper placement within the blocks.

Furthermore, the autocorrelation analysis of transaction failure rates demonstrates statistically significant temporal dependencies, indicating that recent transaction failure rates are influenced by preceding ones. The analysis also identifies recurring patterns at regular intervals, most prominently at 24 hours, suggesting the consistent reoccurrence of specific operational processes or user behaviors within a structured daily cycle.

\rqbox{
\textbf{Finding 3:} 
Failure rates exhibit a strong positive correlation with the volume of failed transactions on an hourly basis, and demonstrate cyclical patterns at 24-hour interval, indicating structured, recurring patterns in operational processes and user behavior over time.
}

\subsubsection{Micro-Level Analysis}
We then performed a detailed analysis of failed transactions on Solana across four dimensions: their positions in blocks, the transaction fees incurred, the compute units (CU) utilized, and cost efficiency, defined as the transaction fees per CU.

\noindent\textbf{Positions in Blocks.} Failed transactions occupy a median position of 592 (mean: 754, min: 0, max: 8,134) in blocks, whereas successful transactions are positioned closer to the top of blocks, with a median rank of 529 (mean: 764, min: 0, max: 8,264).
A Wilcoxon rank-sum test confirms the statistical significance of the difference in block positions between failed and successful transactions  (p-value < 0.001), suggesting that failed transactions are outcompeted, leading to their deeper placement within the blocks. 

\begin{figure*}[h!]
    \centering
    \vspace{-10pt}
    \begin{subfigure}[b]{0.3\textwidth}
        \centering
        \includegraphics[width=\textwidth]{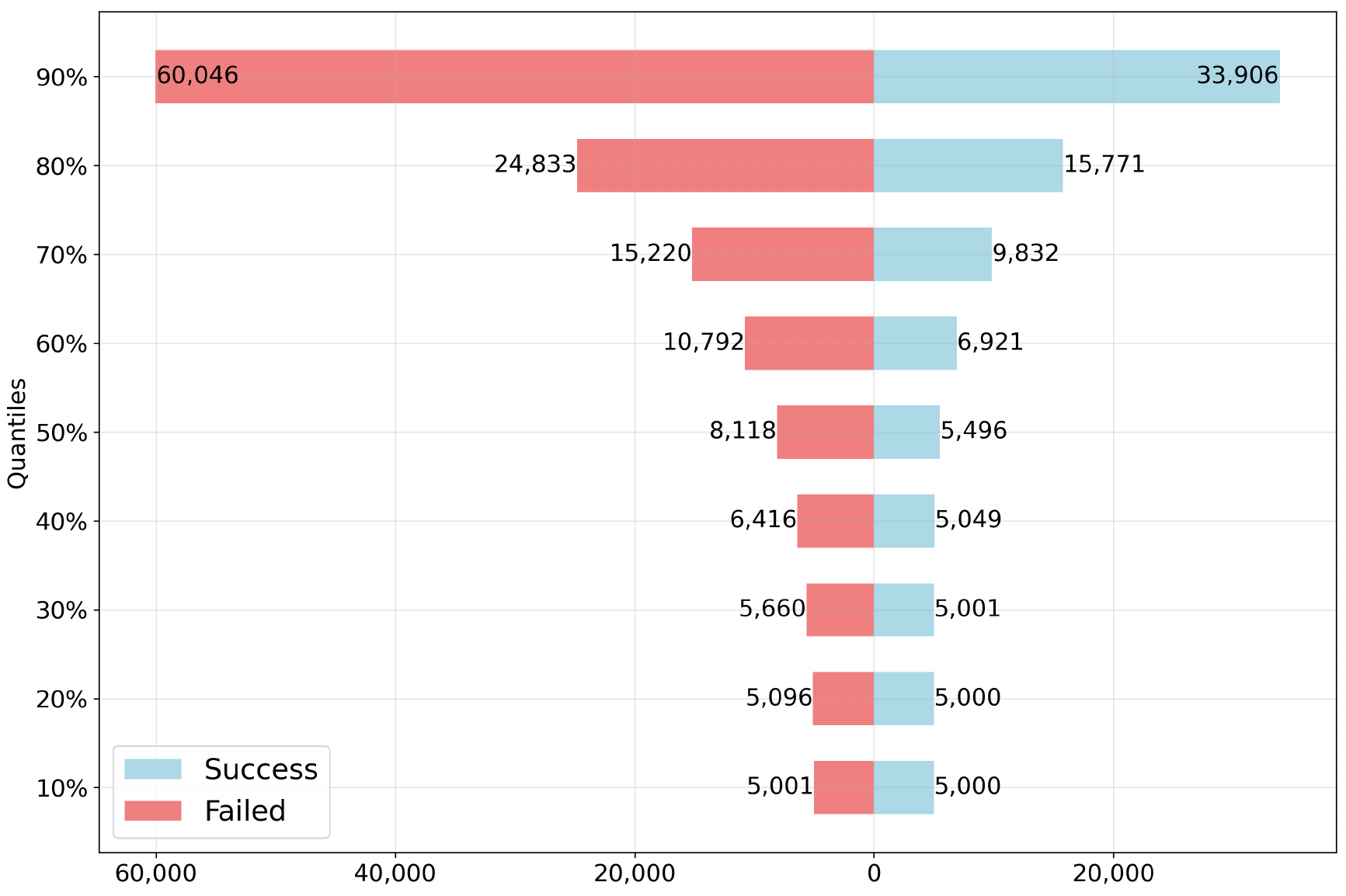}
        \caption{Transaction Fee}
        \label{fig:fee}
    \end{subfigure}
    \hfill
    \begin{subfigure}[b]{0.3\textwidth}
        \centering
        \includegraphics[width=\textwidth]{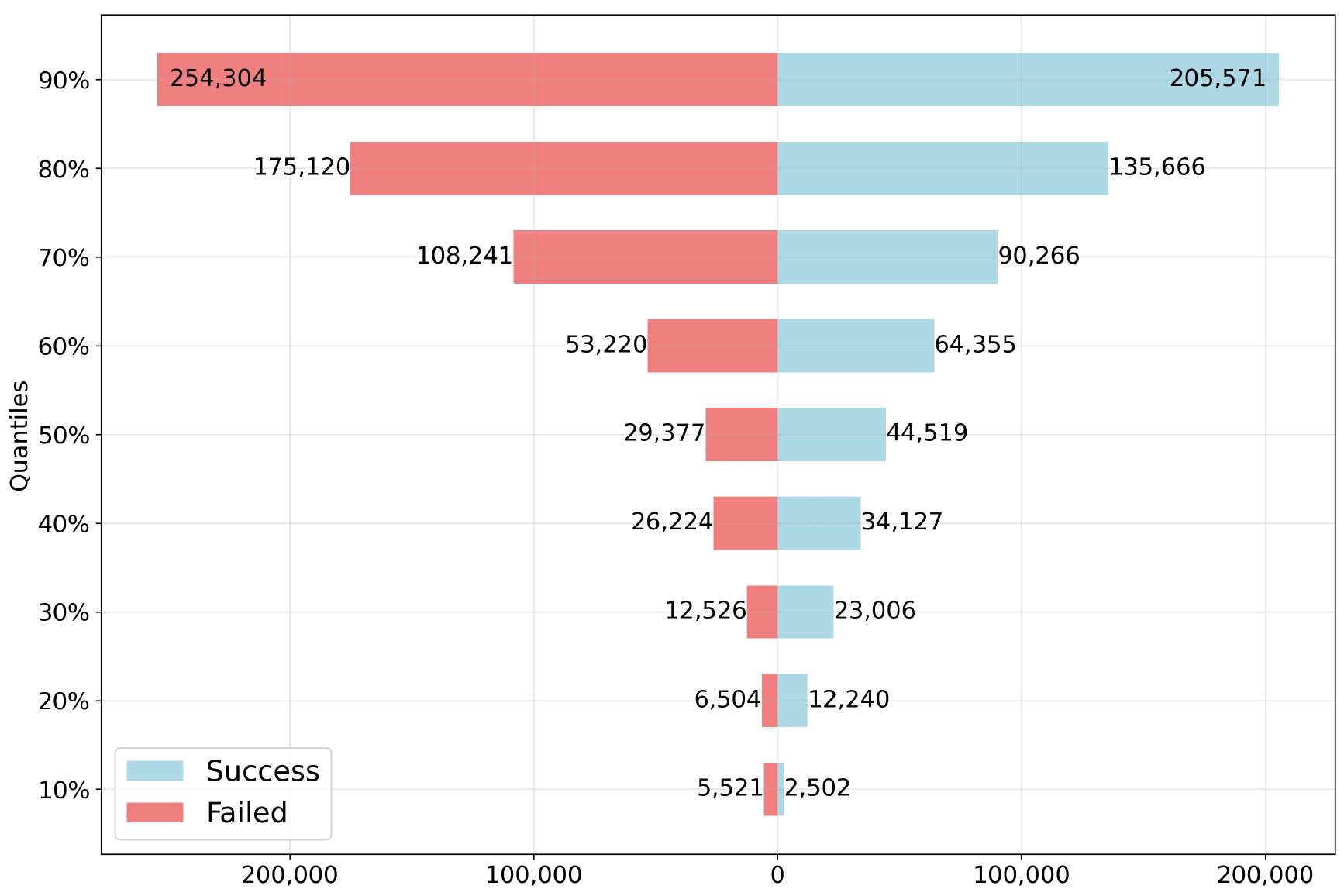}
        \caption{Compute Units (CU)}
        \label{fig:cu}
    \end{subfigure}
    \hfill
    \begin{subfigure}[b]{0.3\textwidth}
        \centering
        \includegraphics[width=\textwidth]{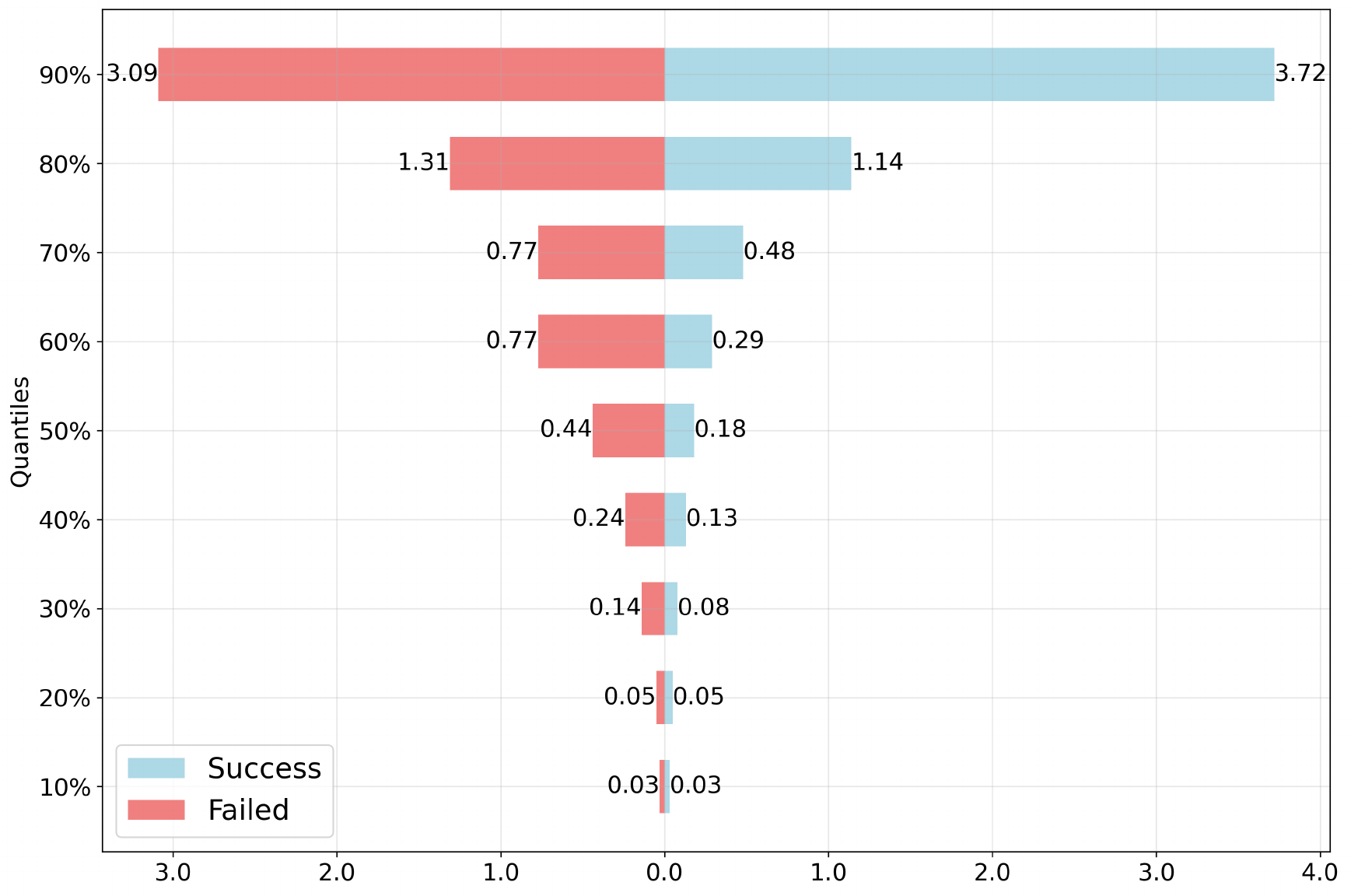}
        \caption{Fee per CU}
        \label{fig:fee_per_cu}
    \end{subfigure}
 \vspace{-10pt}
 \caption{Fees, Compute Units, and Cost Efficiency of Successful vs. Failed Transactions on Solana.}
    \label{fig:tx_cost}
    \vspace{-10pt}  
\end{figure*}

\noindent\textbf{Transaction Fees, Compute Units, and Cost Efficiency.} Figure~\ref{fig:tx_cost} compares successful and failed transactions on Solana across three dimensions: transaction fees, compute units (CUs) consumed, and cost efficiency (fees per CU). 
As shown in Figure~\ref{fig:fee}, failed transactions incur higher transaction fees than successful transactions across all quantiles. Figure~\ref{fig:cu} shows that failed transactions consume fewer compute units than successful transactions, particularly below the 70\% quantile, which may indicate an under-allocation of computing resources~\cite{solanadocpriorityfee}. Meanwhile, as depicted in Figure~\ref{fig:fee_per_cu}, failed transactions exhibit higher fees per CU compared to successful transactions, especially below the 80\% quantile, suggesting potential resource misallocation or suboptimal execution of failed transactions.
The Wilcoxon rank-sum test yielded p-values less than 0.001 for all three dimensions, confirming statistically significant differences between failed and successful transactions in terms of transaction fees, compute units, and cost efficiency.

In summary, the failed transactions incur higher transaction fees, indicating their attempt to prioritize their processing in line with the fee-based prioritization system of Solana. However, the failed transactions have be outcompeted by the successful ones, as indicated by their later positions in blocks. The significantly lower CU and higher fees per CU of failed transactions further suggests the inefficiency in the execution of failed transactions, which may explain why failed transactions have been outcompeted by successful ones. The inefficiency highlights the need for Solana program developers to implement mechanisms that provide users initiating failed transactions with timely feedback, include program-specific guidance on optimizing transaction submission timing, fee estimation, and compute unit allocation, rather than relying on indiscriminate fee increases to improve transaction success rates.

\vspace{-5pt}
\rqbox{
\textbf{Finding 4:} 
Failed transactions tend to appear in later positions in blocks compared to successful ones (median position 592 vs. 529) and exhibit higher transaction fees, as well as higher fees per CU, despite using lower compute units, suggesting a relative inefficiency in the execution of failed transactions. The observed inefficiency highlights the need for Solana program developers to implement mechanisms that provide users initiating failed transactions with timely feedback, include program-specific guidance on optimizing transaction submission timing, fee estimation, and compute unit allocation, rather than relying on indiscriminate fee increases to improve transaction success rates.
}

\section{Errors in Failed Transactions (RQ2)}\label{sec:RQ2}
\subsection{Methodology}
To answer RQ2, we first extracted the error message from each error log entry in the log messages of the failed transactions in our dataset by taking into account their development frameworks through the following steps.
\blackcircle{1} \textbf{\textit{Anchor framework development:}} For the failed transactions originating from the programs that are developed based on the Anchor framework\footnote{\url{https://www.anchor-lang.com/}}, the error log entries follow a specific macro defined in the Anchor framework, i.e., {\tt \#[error(error\_code)]}. These entries are formatted as {\tt Program log: AnchorError occurred. Error Code: <Error\_Code>. Error Number: <Error\_Number>. Error Message: <Error\_Message>}. We extracted such error log entries based on the {\tt Error\_Message} field.
\blackcircle{2} \textbf{\textit{Non-Anchor framework development:}} For the failed transactions originating from the programs that are built with no support from the Anchor framework, we located the corresponding error log entries by searching for the keywords in the log messages, such as \textit{error}, \textit{failed}, \textit{invalid}, \textit{panicked}, and \textit{mismatched}. For example, in \autoref{lst:log}, we identified line 6 as the error log entry and extracted ``exceeds desired slippage limit'' as the error message.
Note that we excluded error logs that contained incomplete, truncated, or missing error messages from failed transactions.
Consequently, we extracted 2,311 unique error messages from 1,145,479,366 failed transactions in our dataset.

Next, we applied thematic analysis~\cite{cruzes2011recommended} to the error messages for classification, through the following steps. 

\noindent\blackcircle{1}~\textbf{\textit{Coding:}} We began by coding the top 100 error messages that are most frequently occurring in the log messages of the failed transactions in our dataset.
For each error message, we randomly selected multiple failed transactions from our dataset that were associated with the error message, until adequate information for determining its code. Two labelers reviewed the complete log messages of the selected failed transactions, as well as the source code of the corresponding programs if available. To ensure the quality of the codes, the two labelers discussed the initial codes to reach an agreement. As a result, 34 codes emerged in the initial round of coding.
We then continued to expand the scope of coding by considering 73 additional error messages with more than 50,000 occurrences, until no new code emerged.
Consequently, we identified 11 additional codes from the error messages, resulting in 45 codes in total, accounting for 99.71\%  of the error messages associated with the failed transactions with explicit error messages of our dataset. 
Among the 173 error messages we coded, 61 were associated with at least one open-source programs, for which the source code was available for reference during the coding process, while 112 were triggered exclusively by closed-source programs. For the 91 error messages that lacked source code, we derived codes based on the literal content of the corresponding log messages. For example, the error message \textit{Slippage tolerance exceeded}, which is associated with 150,672,490 failed transactions originating from the Jupiter Aggregator V6 in our dataset, was coded as ``Slippage''. The remaining 21 error messages, for which no source code was available and whose literal content was ambiguous, were coded as ``Unknown''.

\noindent\blackcircle{2}~\textbf{\textit{Card Sorting:}} 
We used the open card sorting technique~\cite{spencer2009card}, a widely adopted technique for organizing items without predefined categories. Specifically, the two labelers separately sorted the codes into potential themes for thematic
similarity, following a similar process as prior studies, e.g., LaToza et al.'s study~\cite{latoza2006maintaining}. 

Any discrepancies between the two labelers were resolved through discussions with another author of the paper, an expert in Solana, to reach a consensus. For further validation, the third author reviewed the derived codes during the coding phase and classified the codes according to the proposed themes. No misclassification cases were identified in the validation process.

\subsection{Results}
\begin{figure}
    \centering
    \includegraphics[width=0.5\linewidth]{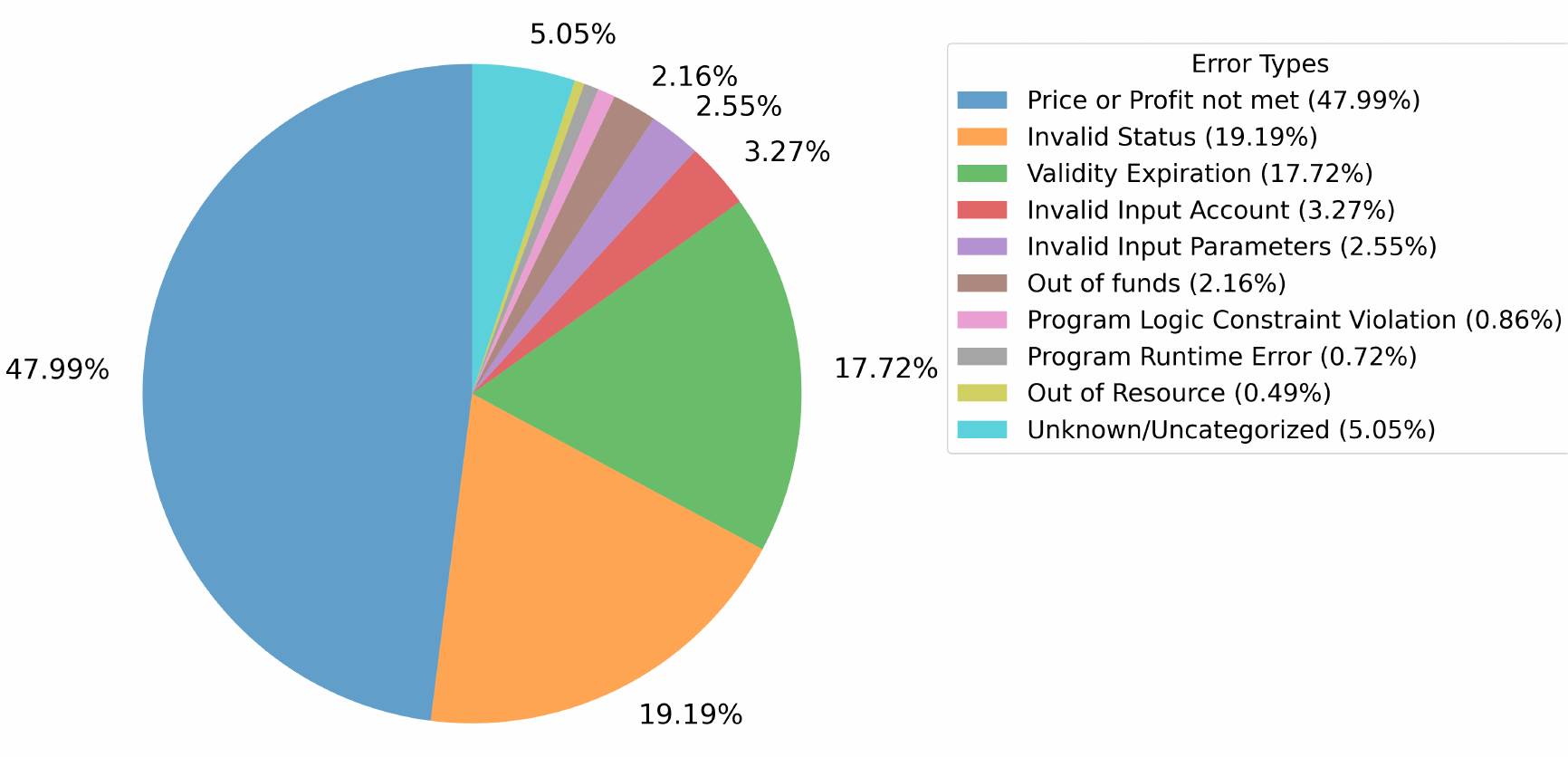}
    \vspace{-10pt}
\caption{Distribution of Error Types in Failed Transactions.}
    \label{fig:error_type}
    \vspace{-15pt}
\end{figure}

Figure~\ref{fig:error_type} illustrates the distribution of the ten error types we identified in failed transactions, revealing a long-tail distribution. The three most prominent error types, \emph{price or profit not met} (47.99\%), \emph{invalid status} (19.19\%), and \emph{validity expiration} (17.72\%), together account for 84.90\% of all failures, indicating the critical role of business logic validation, state management, and infrastructure reliability in ensuring transaction success on Solana. 
Secondary error types, such as \emph{invalid input account} (3.27\%), \emph{invalid input parameters} (2.55\%), and \emph{out of funds} (2.16\%), indicate operational challenges related to resource availability and input validation. 
Less frequent error types, including \emph{program logic constraint violation} (0.86\%), \emph{program runtime error} (0.72\%), and \emph{out of resource} (0.49\%), though rare, highlight the presence of potential corner cases and programming defects that could impact system robustness under specific conditions. 
The following provides a detailed description of each error type, accompanied by the corresponding error logs and representative transaction examples.

\noindent\faExclamationCircle~\textbf{Price or Profit not Met}
serves two essential purposes: (1) protecting users from potential financial losses due to price slippage, and (2) mitigating the risk of front-running attacks, which exploit latency and market inefficiencies to gain unfair advantages~\cite{mev}. The relevant error logs of this type are presented below:
\begin{itemize}[nosep]
    \item[\faTwitch] \emph{Slippage tolerance exceeded:} 
    The error occurs when token prices fluctuate between the submission and processing of a transaction, and a low slippage tolerance causes the transaction to fail if the fluctuation of token prices exceeds the specified threshold.
    A representative example of this error arises from the transaction \texttt{5BB13Yde}~\cite{tx_5BB13Yde}, which involves an attempt at cyclic arbitrage to acquire USDT tokens. The transaction enforces a zero-slippage tolerance, requiring an exact match in exchange rates to proceed. Due to this strict condition, the transaction fails when the final balance of USDT tokens proves insufficient, resulting in the loss of only the transaction fees.
    
     \item[\faTwitch] \emph{IOC order failed to meet minimum fill requirements:} 
     The error log reflects the failure of an Immediate-Or-Cancel (IOC) order to meet the required minimum quantity or price conditions, leading to its automatic cancellation. On the Solana blockchain, IOC orders are often employed in DEXs to ensure efficient execution by either fulfilling the specified conditions immediately or canceling the order to prevent unfavorable outcomes~\cite{solana_ioc_order}. If market conditions shift, such as slippage exceeding the set tolerance or liquidity being insufficient, the order fails to execute within the required parameters. 
\end{itemize}

\noindent\faExclamationCircle~\textbf{Invalid Status} occurs when transaction initiators attempt to execute operations on program-owned data or off-chain data that is not in a valid state to permit the requested operations. The relevant error logs of this type are presented below:
\begin{itemize}[nosep]
    \item[\faTwitch]\emph{The amm account owner is not match with this program:} 
    The error occurs when the on-chain data account of a liquidity pool, which stores essential information such as the state of the pool and its fee rates, is not yet finalized at the time of creation. If a swap transaction is initiated before the completion of the initialization of the pool, the operation will fail with this error. 
    The error is often caused by sniper bots that detect pool creation transactions and attempt to execute trades prematurely, aiming to be among the first transactions processed in the newly created pool. For instance, the scheduled opening time of the BLAST-SOL pool was March 6, 2024, at 08:05:13 UTC. However, the transaction \texttt{5AnmV4S}~\cite{tx_5AnmV4S} attempted to perform a swap 49 seconds prior to the initialization of the pool, leading to the error.

    \item[\faTwitch]\emph{Account is frozen:} 
    The error log indicates that a transaction attempts to interact with a token account that has been frozen by the mint authority, rendering all token operations unavailable. Such freezing may be associated with honeypot attacks, where malicious actors issue fraudulent tokens to attract investments and subsequently freeze the funds of investors, preventing them from making withdrawals~\cite{account_frozen}.

\end{itemize}

\noindent\faExclamationCircle\textbf{~Validity Expiration} 
occurs when a transaction fails to execute within its designated validity window. On Solana, each transaction incorporates a {\tt recent\_blockhash} field to define the time window during which it remains valid. If the transaction reaches a validator after the expiration of its validity window, typically due to network congestion or connectivity disruptions, the validator will reject the transaction, thereby mitigating the risk of replay attacks~\cite{confirmation}. Examples of error logs for this error type include include \faTwitch~\emph{slot expired}, \faTwitch~\emph{stale report}, and \faTwitch~\emph{time expired}.

\noindent\faExclamationCircle\textbf{~Invalid Input Account} 
occurs when a user provides unauthorized or incorrect accounts for an instruction or omits the required accounts. In the stateless programming model of Solana, instructions depend on the transaction initiators to specify the accounts on which the operations will act. If the provided accounts are invalid or do not align with the requirements of the instructions, such error type will be triggered. The relevant error logs of this type are presented below:

\begin{itemize}[nosep]
    \item[\faTwitch]\emph{Not enough account keys given to the instruction:} 
    The error log is generated when the required account addresses are not adequately supplied for the execution of an instruction. For example, the transaction \texttt{ebX6rwkn}~\cite{tx_ebX6rwkn} involves a complex operation consuming 267,105 compute units and interacting with 16 distinct programs. Although 66 input accounts are provided for the \texttt{3J3HFc8} program instruction, the transaction encounters such an error. For operations of this scale, it can be particularly challenging for transaction initiators to identify all required account addresses and arrange them in the correct sequence to ensure successful execution.
    
    \item[\faTwitch]\emph{InvalidSplTokenProgram:} 
    The error log occurs when the owner of the provided input account does not align with the expected address of the Solana native token program. For instance, when the account for a specific token has not been properly initialized, the use of such an invalid address in a token swap operation will trigger such error log~\cite{invaildspl}.
\end{itemize}

\noindent\faExclamationCircle\textbf{~Invalid Input Parameters}
arises when a user provides an argument for an instruction that falls outside the acceptable range, either too small, too large, or in an incorrect format. Such errors often occur because programs impose specific requirements on input data, such as predefined ranges or structures, which transaction initiators may overlook, particularly if the program documentation does not clearly specify these requirements.
An example of error logs for this error type is \faTwitch~\textit{InvalidInput}, which indicates that the provided instruction input is invalid. For instance, the Raydium program enforces that the minimum output swap amount must be greater than zero. However, in the transaction \texttt{zxFnAz}~\cite{tx_zxFnAz}, the input parameter for the minimum swap output is incorrectly set to 0, triggering this error log.

\noindent\faExclamationCircle~\textbf{Out of Funds} 
indicates that Solana users lack sufficient tokens in their wallets to cover the transaction amount along with the associated fees. The Raydium program enforces a minimum balance of SOL in the wallet to facilitate trades or swaps~\cite{raydium_out_of_funds}. Transaction initiators may fail to account for these fees, leading to this error type.
Examples of error logs for this error type include \faTwitch~\textit{insufficient funds}, \faTwitch~\textit{insufficient collateral}, and \faTwitch~\textit{insufficient funds for instruction}.

\noindent\faExclamationCircle~\textbf{Program Logic Constraint Violation} occurs when a transaction violates program logic constraints and cannot be executed. Examples of error logs for this error type include \faTwitch~\emph{Player has not requested a result} and \faTwitch~\emph{A seeds constraint was violated}. 
It is noteworthy that infrequent failures (fewer than 0.05 million failed transactions) are associated with the error message \faTwitch~\emph{The provided victim signature was already used}, as exemplified by a failed transaction on December 14, 2023~\cite{tx_3HdLQW}. Such failures may indicate the potential occurrence of security attacks, which could serve as a critical signal for program developers to investigate the integrity of the associated programs.

\noindent\faExclamationCircle~\textbf{Program Runtime Error} arises when a transaction triggers an unhandled corner case that causes the program to abort or panic. Examples of error logs for this error type include \faTwitch~\emph{panicked at range end index 72 out of range for slice of length 0}, \faTwitch~\emph{panicked at called {\tt Option::unwrap()} on a {\tt None} value}, and \faTwitch~\emph{panicked at attempt to subtract with overflow}.

\noindent\faExclamationCircle~\textbf{Out of Resource} 
arises when a transaction exhausts its allocated resources, such as compute units, or exceeds the runtime constraints imposed by Solana, including constraints on heap memory and cross-program invocation data size~\cite{out_of_memroy}. Examples of error logs for this error type include \faTwitch~\emph{memory allocation failed, out of memory}, \faTwitch~\emph{computational budget exceeded}, and \faTwitch~\emph{exceeded CUs meter at BPF instruction}. For instance, the transaction \texttt{26hZfe}~\cite{tx_26hZfe} failed due to memory exhaustion while interacting with 93 accounts via Jupiter Aggregator V6. 

\rqbox{
\textbf{Finding 5:}  
We identify ten distinct error types from the error messages of the failed transactions on Solana. The distribution of these error types across the failed transactions follows a long-tail pattern, with the top three types, \textit{price or profit not met}, \textit{validity expiration}, and \textit{invalid status}, accounting for 47.99\%, 19.19\%, 17.72\% of transaction failures, respectively, which indicate challenges in business logic validation, state management, and infrastructure reliability on Solana. Other error types, such as \textit{out of funds} and \textit{invalid input parameters}, suggest potential operational challenges related to resource availability, user input accuracy in the Solana ecosystem.}

\section{Contribution of Programs and Accounts to Errors (RQ3)}\label{sec:RQ3}
\subsection{Methodology}
To answer RQ3, we first identified the programs with the highest volume of failed transactions and categorized their failures according to the error types established in RQ2, revealing the specific error types associated with each program. Next, we analyzed the distributions of error types associated with two account types, and performed a comparative analysis of the error distributions between bot-driven and human-driven transactions. Moreover, we investigated how program functionalities and the distinct behaviors of these initiators contribute to the emergence of specific error types, aiming to provide insights into the root causes underlying transaction failures.

\subsection{Results}

\begin{figure*}
    \centering
    \begin{subfigure}[b]{0.33\textwidth}
        \centering
        \includegraphics[width=\linewidth]{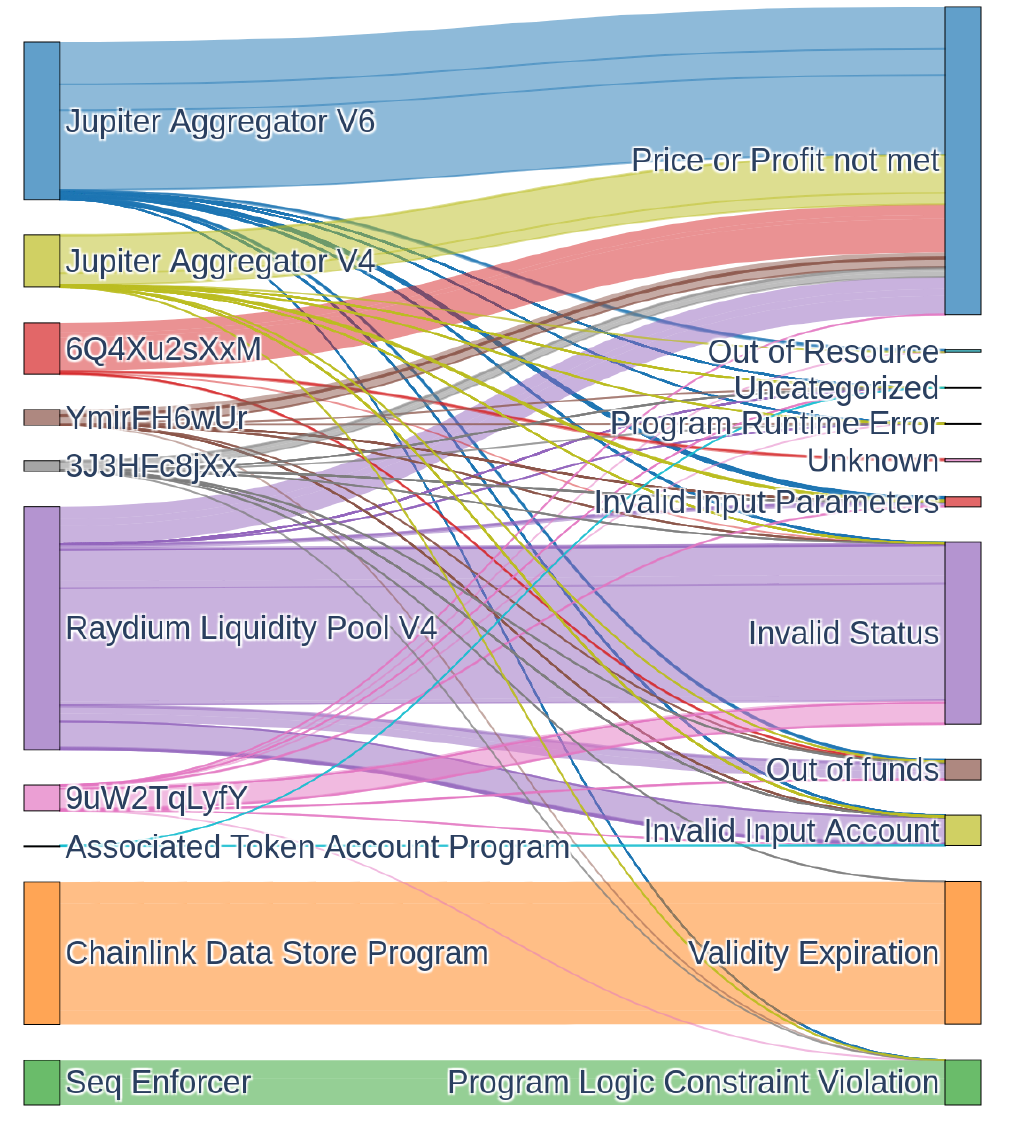}
        \caption{Top 10 Programs}
        \label{fig:program_sankey}
    \end{subfigure}
    \hfill
    \begin{subfigure}[b]{0.33\textwidth}
        \centering
        \includegraphics[width=\linewidth]{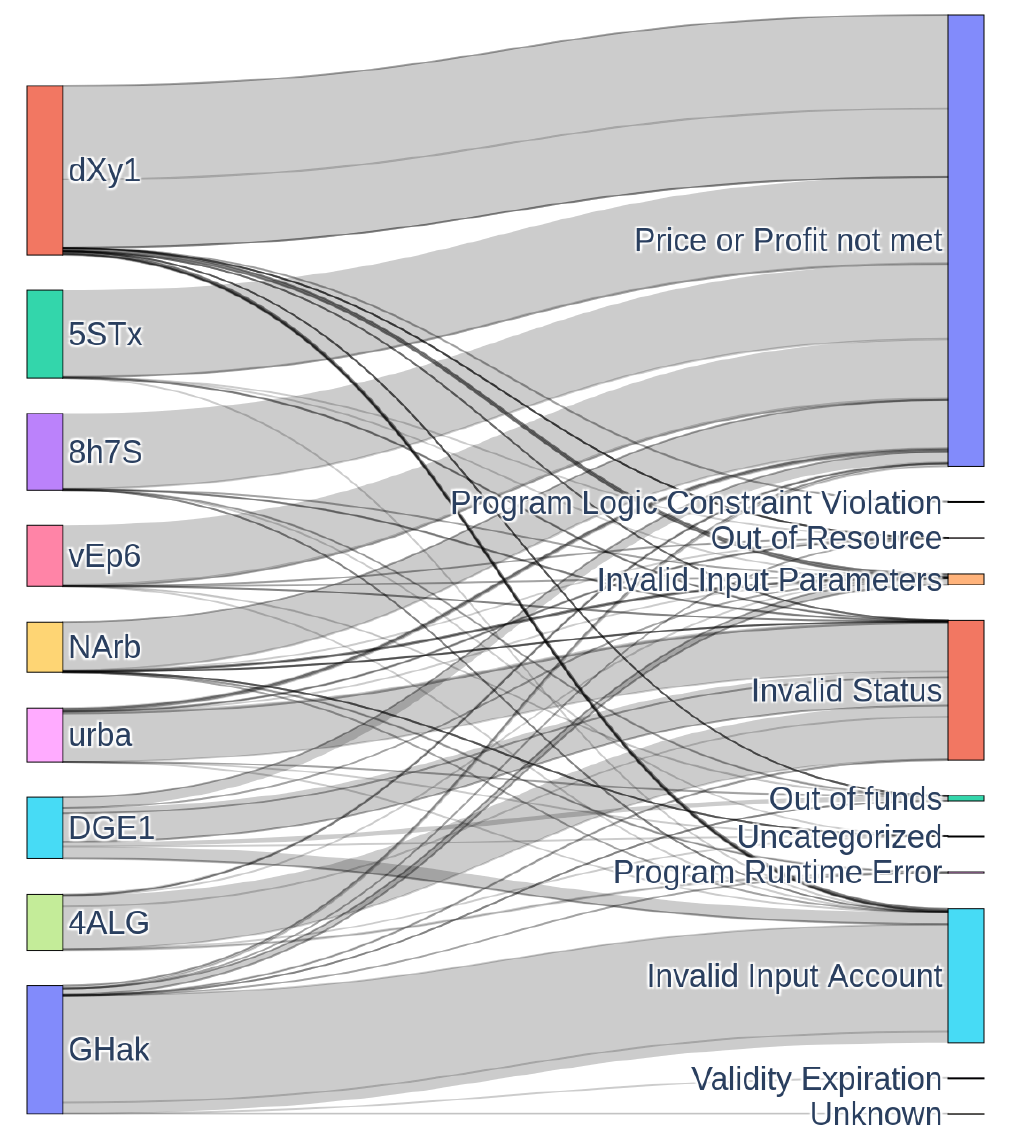}
        \caption{Top 10 Bots}
        \label{fig:bot_sankey}
    \end{subfigure}
    \hfill
    \begin{subfigure}[b]{0.32\textwidth}
        \centering
        \includegraphics[width=\linewidth]{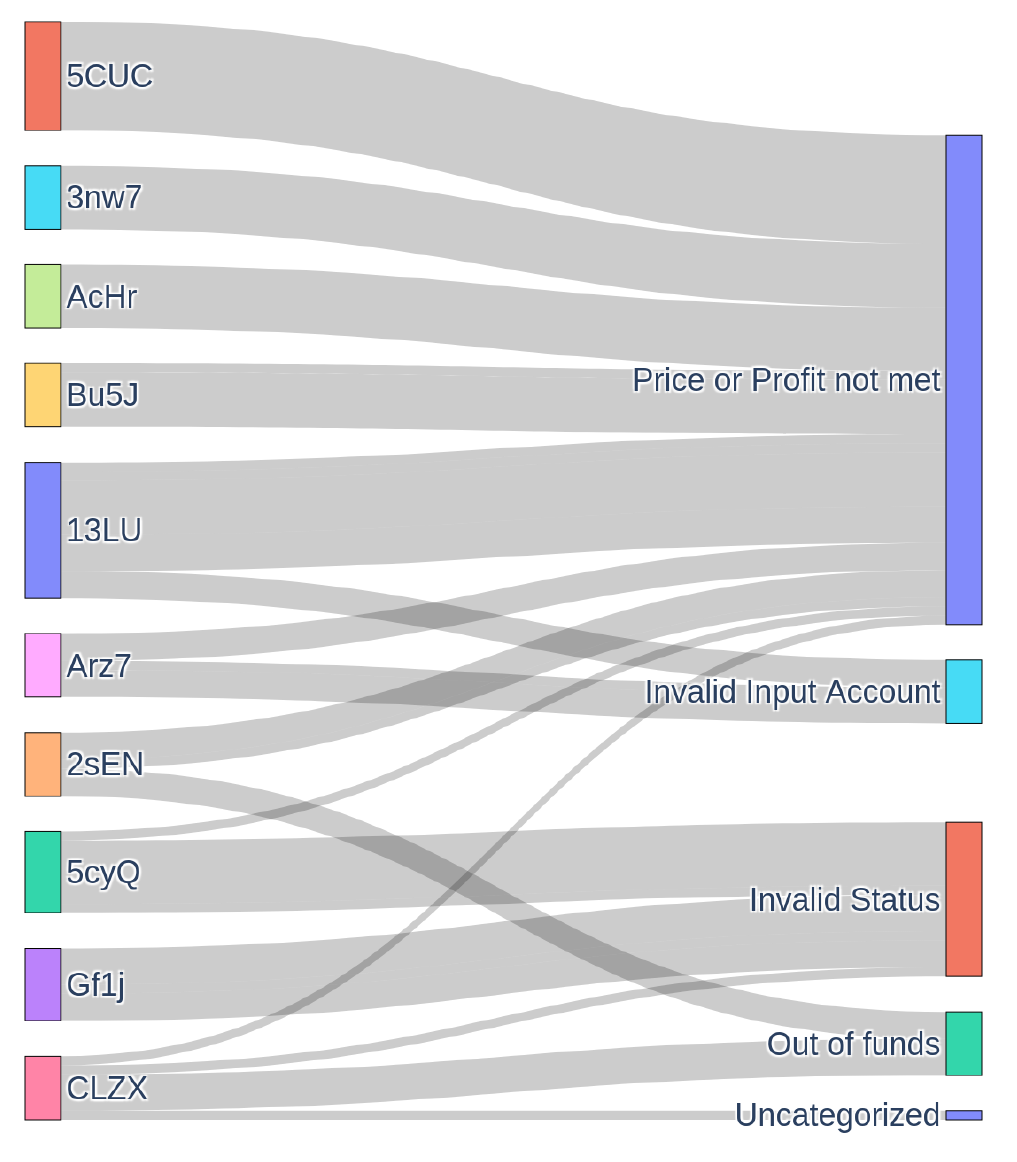}
        \caption{Top 10 Human Accounts}
        \label{fig:human_sankey}
    \end{subfigure}
    \vspace{-10pt}
\caption{Mappings of Error Types Across Top 10 Programs, Bots, and Human Accounts with the Highest Volumes of Failed Transactions.}
    \label{fig:error_rq3}
    \vspace{-20pt}
\end{figure*}

Fig.~\ref{fig:program_sankey} illustrates the distribution of various error types across different programs with top 10  failed transactions on the Solana blockchain. Each program is listed on the left side, with connections flowing to specific error types on the right, highlighting how different programs contribute to various error types of transaction failures. We made the following observations:

\begin{itemize}[nosep]
    \item Jupiter Aggregator V6 and Jupiter Aggregator V4 contribute significantly to the \emph{price or profit not met} errors. The two programs aggregate DEXs and facilitate financial swaps, where pricing mismatches or arbitrage conditions not being met are common causes of transaction failure. 
    \item Raydium Liquidity Pool V4  primarily causes the \emph{invalid status} error. As suggested by a representative error message ``The amm account owner does not match with this program'', the program tends to be frequently targeted by sniping bots or misused, leading to failed status checks.

    \item The Chainlink Data Store program is identified as a significant contributor to the \emph{validity expiration} error. 
    Specifically, as indicated by the frequently occurred error message ``Stale report'', Chainlink seeks to notify the initiators of the failed transactions that the data provided by Chainlink is subject to expiration~\cite{chainlink_intro}.

    \item The two private programs, 6Q4Xu2sXxM and 9uW2TqLyfY, primarily trigger the \emph{price or profit not met} and \emph{invalid status} errors, aligning with behaviors seen in arbitrage bots and snipping bots, respectively, and consistent with the observations in Reddit discussions~\cite{prg_6qex, prg_quw, peppermints}.
\end{itemize}

\rqbox{
\textbf{Finding 6:}  
DEX aggregators frequently encounter failures attributed to \textit{price or profit not met} errors, while AMMs are primarily impacted by \textit{invalid status} errors and tend to be targeted by sniper bots. Additionally, private programs demonstrate error patterns indicative of bot-like activity.}

Fig.~\ref{fig:bot_sankey} and Fig.~\ref{fig:human_sankey} present the distributions of error types across the accounts of two distinct categories, bot accounts (b) and human accounts (c), with top 10 failed transactions. In each figure, each account is listed on the left side, with connections flowing to specific error types on the right,
highlighting how different accounts contribute to various error types of transaction failures. We made the following observations:

\begin{itemize}
    \item[\faRobot~\faUser] The \emph{price or profit not met} error accounts for 60.65\% and 63.95\% of the top 10 bot-initiated and top 10 human-initiated failed transactions, respectively, suggesting the frequent potential interactions of trading bots and human users with AMMs.
    \item[\faRobot] A significant number (18.78\%) of the top 10 bot-initiated failed transactions result in the \emph{invalid status}  error, which may be due to the fact that bots frequently engage with liquidity pools or other stateful programs, thus leading to status mismatches. For instance,  as the frequent occurrence of the error log ``The amm account owner does not match with this program'' suggests, bots tend to interact with outdated or unapproved accounts, indicating misuse or aggressive strategies of sniping bots.
    \item[\faRobot] The \emph{invalid input account} error accounts for a large number of the failed transactions (17.80\%) initiated by the top-10 bot-initiated failed transactions, which dominates the failed transactions of the GHak bot, suggesting the tendency of the bot in providing incorrect account addresses as parameters when initiating transactions.
     \item[\faUser] The \emph{invalid status} error accounts for 19.77\% of the top 10 human-initiated failed transactions. For instance, the error log ``Account is frozen'' suggests that users may have invested in potentially fraudulent tokens, leading to the freezing of the associated token accounts in order to prevent further user losses.
     \item[\faUser] The \emph{out of funds} error accounts for 8.14\% of the top 10 human-initiated failed transactions, suggesting that human users failed to provide adequate fees when initiating transactions.
\end{itemize}

In comparison, bot accounts trigger a broader range of error types as compared to human accounts. The difference in the distributions of error types between human and bot accounts suggests that bots tend to face more complicated challenges due to their high transaction volume and interaction with more advanced financial operations in the Solana ecosystem, as compared to human users.

\rqbox{
\textbf{Finding 7:}  
Bot accounts exhibit a broader range of error types in their failed transactions as compared to human accounts. The \emph{price or profit not met} error accounts for the majority of the failed transactions for both bot and human accounts.
Human-initiated transactions are
more prone to \emph{out of funds} errors as compared to bot-initiated transactions.
The difference in the distributions of error types between bot and human accounts indicates that bots encounter more complex challenges arising from their high transaction volumes and interactions with advanced operations in the Solana ecosystem.
}

\section{Discussion}\label{sec:discussion}
\subsection{Comparison across Blockchain Ecosystems}
\noindent\textbf{Failure Rate.}
We observed a transaction failure rate of 6.22\% for human accounts on Solana (Finding 1), which is comparable to the transaction failure rate ranging from 1\% to 3\% on Ethereum, as reported in previous studies~\cite{oliveira2021analyzing,eth_failed_rate}. In contrast, bot accounts on Solana exhibited a significantly higher transaction failure rate of 58.43\% (Finding 1).  The observed discrepancy in transaction failure rates for bot accounts between Solana and Ethereum can be ascribed to the fundamental architectural differences between the two platforms. Specifically, Solana's parallel processing model of transactions and low transaction fees facilitate high throughput, but may unintentionally encourage bot activity. In contrast, Ethereum's sequential processing model of transactions and gas auction mechanism inherently constrain the volume of bot activity.
As prominent Layer 2 scaling solutions for Ethereum, Base and Arbitrum employ rollups to bundle multiple transactions and execute them off-chain, thereby alleviating network congestion and reducing transaction fees on Ethereum~\cite{eth_layer2,arbitrum}. The reduction in transaction fees on Base and Arbitrum incentivizes an increase in transaction volume, resulting in observed transaction failure rates of 21\% and 15.4\% on Base and Arbitrum, respectively~\cite{eth_layer2_failure}.  Nonetheless, the failure rates on Base and Arbitrum remain significantly lower than the 58.43\% observed for bot accounts on Solana (Finding 1).

\noindent\textbf{Failure Patterns.} 
We identified ten distinct error types from the error messages of failed transactions on Solana (Finding 5), three of which are also reported in the failed transactions of Ethereum~\cite{eth_failed_tx}, namely \emph{Out of Funds (Gas)}, \emph{Program (Contract) Runtime Error}, and \emph{Invalid Status}.
The most frequently encountered error type in failed transactions varies between Solana and Ethereum. Specifically, \textit{Price or Profit Not Met} accounts for the majority of failed transactions on Solana (47.99\%) (Finding 5), while \emph{Out of Funds (Gas)} is the leading error type of failed transactions on Ethereum~\cite{eth_failed_tx_3, eth_failed_tx_2}.
The difference in the dominant error type of failed transactions indicates that Solana's failures are predominantly attributed to non-profitable transactions in a high-volume, low-cost platform, while Ethereum's failures primarily result from insufficient gas for transaction execution in a high-cost, congested platform. The difference also reflects the differing trade-offs between transaction throughput and cost predictability inherent in the design of the two platforms.

Due to the limited availability of public transaction failure data for Bitcoin and Binance Smart Chain (BSC), we suggest that future work could put efforts into collecting failure data from the Bitcoin and BSC ecosystems, to enable comprehensive comparisons of transaction failures across different blockchain ecosystems.

\subsection{Implications}
\noindent\textbf{Ecosystem-wide strategies could be beneficial in reducing the elevated rate of failed transactions on Solana.}
We observe a substantial incidence of failed transactions on Solana, possibly exacerbated by bots repeatedly submitting transactions, which further contributes to potential network congestion on Solana (Finding 1). 
We also note a correlation between increased transaction surges and elevated failure rates on the network (Finding 3).
To prevent bots from overwhelming the network and to enhance user experience, Solana could benefit from implementing comprehensive, ecosystem-level strategies.
Ethereum, by contrast, leverages a gas price auction mechanism that effectively curtails bot activity through elevated transaction fees during periods of congestion~\cite{faqir2021effect,spain2020impact}. 
However, the gas price auction mechanism on Ethereum can introduce significant volatility in gas prices, particularly during periods of network congestion, leading to unpredictable transaction costs and a poor experience for users with limited resources. In light of the trade-offs associated with the gas auction mechanism on Ethereum, the Solana ecosystem could consider incorporating a novel dynamic fee structure that automatically adjusts the costs for rapid, repeated transaction submissions from the same account. Such a dynamic fee structure could help moderate bot activity while preserving the low transaction fees of Solana, potentially balancing network accessibility with improved resilience by mitigating transaction failures.
Future work could involve a comparative analysis of transaction fee designs across blockchain ecosystems, with a focus on evaluating the potential impact of varying transaction fee designs on both the network accessibility and overall resilience of the Solana ecosystem.

\noindent\textbf{DeFi protocol mechanisms are necessary to mitigate bot-induced transaction failures on Solana.}
The high frequency of \textit{invalid status} errors observed in Raydium suggests that this AMM has become a primary target for sniper bots on Solana (Finding 6). Engaging in speculative front-running within newly launched liquidity pools, these sniper bots contribute to a substantial volume of failed transactions, potentially gaining disproportionate market advantages and introducing risks to market stability~\cite{cernera2023sniper}. To address this challenge, developers of AMMs on Solana might consider adopting mechanisms similar to those on Ethereum, such as imposing elevated transaction fees or restricting token purchases per transaction and/or per address during the initial phase of a liquidity pool~\cite{cernera2023sniper}.
On the other hand, another prevalent error type, \textit{price or profit not met}, is commonly associated with arbitrage bot activity directed at Jupiter Aggregator V6, a DEX aggregator (Finding 5 and Finding 6). 
The error type indicates that arbitrage bots cause transaction failures when anticipated profits do not materialize or market conditions change during execution, likely due to the aggressive profit-seeking strategies they employ. To mitigate such transaction failures, DEX aggregators could implement differentiated aggregator service fee structures, where transactions initiated by arbitrage bots are subject to higher fees than those initiated by human users. The differentiated aggregator service fee structures could help disincentivize arbitrage bots from engaging in excessive profit-driven activities, thereby reducing the occurrence of transaction failures.

\noindent\textbf{The need for automated tools to address development challenges of human accounts on Solana.}
Human accounts frequently encounter \textit{out of funds} errors, indicating the necessity for accurate token cost calculation to prevent wallet depletion (Finding 7). Automated token cost calculation tools would assist users in managing balances more effectively, thereby reducing such errors.
Besides, failed transactions appear in deeper block positions with higher per-unit fees (Finding 4). The development of precise computation unit estimation tools and gas fee calculators would help users optimize transaction placement and increase success rates.
The frequent occurrence of \textit{invalid input account} and \textit{invalid input parameters} errors indicates a susceptibility among human accounts to input invalidity, particularly in complex transactions (Finding 5). Tools for online account query and account data structure visualization could help Solana developers verify input accounts and evaluate account states, thus reducing input-related failures.
For issues such as \textit{program runtime error}, introducing automatic testing and debugging tools for Solana programs could improve program reliability and help prevent runtime exceptions in transactions.
In addition, the error type of \textit{account is frozen}, which may indicate potential attackers on Solana, could be leveraged in the future development of intrusion detection systems. For instance, transactions associated with this error type could be integrated with other behavioral data, serving as a multi-dimensional training dataset, thereby enhancing the ability of intrusion detection systems to detect and mitigate sophisticated attacks on Solana.

\subsection{Threats to Validity}
\noindent\textbf{Internal Validity.}
The classification of error messages for failed transactions involves manual inspection, which may introduce potential threats to internal validity due to subjective interpretations by labelers. To mitigate such threats, two labelers independently coded the error messages, engaging in regular comparisons and discussions to ensure coding consistency. For instances of disagreement, we facilitated in-depth discussions involving a moderator and consulted literature and contextual data to achieve accurate and reasonable categorization.
For error messages triggered solely by closed-source programs, we assigned codes based on the literal content of the error messages during thematic analysis. To minimize misalignment between the actual error logic of closed-source programs and the literal content of their error messages, we consulted relevant documentation and official websites of the closed-source programs for further validation. To reduce bias from subjectivity, the results of thematic analysis were cross-validated by three of the authors.

\noindent\textbf{Construct Validity.}
We used a classification algorithm to distinguish between bot and human accounts, which may introduce potential threats to construct validity due to potential account misclassifications. To mitigate such threats, we rigorously defined the behavioral characteristics that differentiate bots from human accounts,  based on established literature~\cite{huang2020eosio, niedermayer2024bots}, ensuring clear and well-grounded criteria for classification. 
Moreover, we validated a random sample from the classification results, confirming the reliability of our classifications.

\section{Related Work}\label{sec:related}
\subsection{Studies on Solana}
Several researchers have explored the performance and scalability of the Solana blockchain platform. Li et al.~\cite{li2021bitcoin} examined the evolution of blockchain technologies from Bitcoin to Solana, identifying key scalability and performance challenges in the enterprise adoption of blockchain solutions. Duffy et al.~\cite{duffy2021IoT} investigated the high transaction throughput of Solana, emphasizing its potential to support large-scale IoT business applications. Pierro et al.~\cite{pierro2022can} evaluated the transaction throughput on Solana and user fees for transaction confirmation, highlighting the significantly lower user fees on Solana compared to those on other blockchain platforms.
Other researchers have developed tools to detect vulnerabilities in Solana smart contracts. VRust~\cite{vrust} employs static analysis to detect vulnerabilities in the source code of Solana smart contracts. 
FuzzDelSol~\cite{smolka2023fuzz} utilizes coverage-guided fuzzing to detect vulnerabilities in closed-source, binary smart contracts on Solana.

Previous studies have predominantly explored the performance and scalability of the Solana blockchain, as well as vulnerabilities of Solana programs, yet lack empirical, transaction-based analysis of the ecosystem. Our study addresses that gap by providing the first comprehensive, large-scale exploration of the Solana ecosystem, with a particular emphasis on failed transactions.

\subsection{Studies on Blockchain Ecosystems}
A breadth of prior work has taken blockchain transactions as the primary subject of research. Some studies have applied a range of graph analysis techniques to explore transactions on various blockchain platforms, addressing a variety of aspects, such as anonymity, transaction network and unusual behaviors of Bitcoin~\cite{fleder2015bitcoin,tao2021complex,di2017analysis}, money transfer patterns and smart contracts of Ethereum~\cite{chen2020understanding,chan2017ethereum,oliva2020exploratory}, and the security issues with EOSIO~\cite{huang2020eosio}. 
Meanwhile, other studies have focused on transactions with specific attributes, such as the empirical investigation of private transactions on Ethereum~\cite{lyu2022empirical}, the examination of the relationship between transaction processing times and gas prices on Ethereum~\cite{pacheco2023my}, the prediction of transaction statuses (confirmation or failure) using machine learning models on Ethereum~\cite{oliveira2021analyzing}, and the investigation of concurrency-related transaction failures on Hyperledger Fabric~\cite{chacko2021my}.
In contrast to prior work, our study focuses on the Solana ecosystem, providing a multifaceted characterization of failed transactions on Solana, and evaluating their financial impacts in the Solana ecosystem.

Another spectrum of prior work focuses on the analysis and detection of bot-controlled accounts on various blockchain platforms.
Some studies examine the behaviors of bots on Ethereum and Binance Smart Chain, including token sniping~\cite{cernera2023sniper}, front-running~\cite{qin2022quantifying, daian2020flash, li2023towards}, and sandwich attacks~\cite{qin2022quantifying,zhou2021high}. 
Other research efforts propose techniques for identifying bot accounts on various blockchain platforms, such as EOSIO~\cite{huang2020eosio} and  Ethereum~\cite{niedermayer2024bots,jin2022detecting}. 
Building upon existing bot detection techniques for other blockchain platforms, our study proposes an approach to identify bots on Solana, and provide the first comprehensive characterization of their behaviors.

\section{Conclusion and Future Work}\label{sec:conclusion}
In this paper, we present the first large-scale empirical study of failed transactions on the Solana blockchain, using a curated dataset of over 1.5 billion failed transactions across more than 72 million blocks. Through the dataset we curated, we systematically analyze the characteristics of failed transactions, identify ten distinct error types arising in the failed transactions, and examine the associations of error types with specific programs and transaction initiators, thereby providing insights into the Solana ecosystem.
Future work could put efforts into improving the reliability and usability of the Solana blockchain through advanced error handling and recovery mechanisms, optimization of protocol and program design patterns, and the development of automated testing tools for programs and real-time monitoring for transaction failures.
Moreover, cross-platform empirical studies could provide insights that improve system resilience and user experience, both on Solana as well as other blockchain ecosystems.

\section{Data Availability}
Our replication package is available online: \url{https://github.com/ZXXYy/Solana_Failed_Tx}.

% \begin{acks}
% This research was supported by the National Science Foundation of China (No. 62472383 and No. 62102358), and the Open Research Fund of the State Key Laboratory of Blockchain and Data Security, Zhejiang University.
% \end{acks}

\bibliographystyle{ACM-Reference-Format}
\bibliography{solana}

\end{document}